\documentclass[3p, authoryear]{elsarticle}
\usepackage{amsmath,amssymb,mathrsfs}
\usepackage{stmaryrd}
\usepackage{latexsym}
\usepackage{CJK}
\usepackage{graphicx}
\usepackage{indentfirst}
\usepackage{listings}
\usepackage{xcolor}
\usepackage{booktabs}
\usepackage{stmaryrd}
\usepackage{multirow}
\usepackage{verbatim}
\usepackage{makecell}
\usepackage{lineno,hyperref}
\usepackage{longtable}
\usepackage{hyperref}
\usepackage{upgreek}
\usepackage{mathtools}
\usepackage{enumitem}
\usepackage{float}

\newcommand{\txc}{\text{c}}

\newcommand{\txw}{\text{w}}

\newcommand{\txI}{\text{I}}
\newcommand{\txp}{\text{p}}

\newcommand{\usgn}{\mathrm{sgn}}

\hypersetup{colorlinks,breaklinks,           linkcolor=blue,urlcolor=blue,anchorcolor=blue,citecolor=blue}

\begin{document}
\title{A Mesoscale Model for Interface-Mediated Plasticity: Investigation of Ductile and Brittle Fracture}
\author[cityu,ucla]{Jinxin Yu}
\author[hku,birmingham]{Alfonso H. W. Ngan}
\author[hku]{David J. Srolovitz\corref{cor2}}
\author[cityu]{Jian Han\corref{cor1}}

\address[cityu]{Department of Materials Science and Engineering, City University of Hong Kong, Hong Kong SAR, China}
\address[hku]{Department of Mechanical Engineering, The University of Hong Kong, Pokfulam Road, Hong Kong SAR, China}
\address[ucla]{Department of Materials Science and Engineering, University of California, Los Angeles, California, United States}
\address[birmingham]{School of Metallurgy and Materials, University of Birmingham, Edgbaston, Birmingham B15 2TT, United Kingdom}
\cortext[cor1]{jianhan@cityu.edu.hk}
\cortext[cor2]{srol@hku.hk}
\date{\today}

\begin{abstract}
The presence of interfaces and grain boundaries significantly impacts the mechanical properties of materials, particularly when dealing with micro- or nano-scale samples. 
Distinct interactions between dislocations and grain boundaries can lead to entirely different overall plastic deformation characteristics. 
This paper employs a two-dimensional continuum dislocation dynamics model to investigate the mechanical properties of materials. 
To accurately depict the physical interactions between lattice dislocations and interfaces/grain boundaries, we apply a mesoscale interface boundary condition, considering various numerical cases, such as single and multiple slip systems. 
Apart from affecting strength and ductility, the accumulation of dislocations near the interface can induce local high stress, potentially resulting in brittle fracture near the interface. 
Consequently, materials experience a competition between ductile and brittle fracture modes during the loading process. 
A phenomenon known as the strain hardening rate up-turn is also investigated under different reaction constants, which can be explained by the competition between the rates of dislocation accumulation and reaction at the interface.
\end{abstract}

\maketitle
\section{Introduction}
Most natural crystalline materials and those of  technological interest  are polycrystalline, i.e., composed of single crystal grains of different crystallographic orientations separated/delimited by grain boundaries (GBs). 
The mechanical properties of polycrystals, including their mechanical strength, ductility, and  fracture behavior, are strongly influenced by the properties and spatial arrangement of these GBs (in addition to the intrinsic properties of the  crystals)~(e.g., see \cite{hirth1972influence,kheradmand2010investigation,zhu2023heterostructured}).
In metallic systems, the well-known and widely applicable Hall-Petch relation posits that reducing the mean grain size  leads to increased  strength (even approaching the theoretical strength)~\citep{hall1951deformation,javaid2021dislocation}. 
However, there is also a well-known trade-off with ductility; reducing grain size usually results in decreased ductility. 
In recent years, heterostructured materials have shown promise in achieving improved combinations of strength and ductility, attributed to hetero-deformation induced (HDI) strengthening~\citep{ji2023recent,zhu2021heterostructured}. 
This suggests that  designing appropriate interfaces  and/or optimizing their spatial distribution may provide a viable route to overcome this strength-ductility trade-off. 
In particular, the key feature is the nature of the interactions between dislocations and interfaces. 
Extensive {\it in situ} experiments have revealed that the interactions between lattice dislocations and interfaces are complex and go beyond simple dislocation blocking~\citep{kacher2012quasi,kacher2014dislocation,kacher2014situ}. 
Since dislocations are the carriers of plastic deformation, these interactions directly affect the plastic deformation of materials, subsequently impacting their strength and ductility.

When a dislocation approaches a GB or any other interface, it may undergo some form of reaction in which it decomposes into line defects with different Burgers vectors or react with existing defects. 
These defects may be other dislocations or disconnections in the GB/interface. 
The overall result is that some dislocations may be transmitted into the adjacent grain, some are absorbed by the GB, some are reflected back into the original grain and some are blocked. 
We proposed a mesoscale interface boundary condition (BC)  to describe the interactions between dislocations and interfaces~\citep{yu2025mesoscale}. 
This model captures the dislocation reaction processes, where dislocations from different slip systems interact at the interface while conserving Burgers vectors.
In the framework of the proposed interface BC model, the reaction constant serves as a measure of the rate at which dislocation reactions occur at the interface. 
Notably, the reaction constant depends on the detailed bicrystallography and atomic structure of the interfaces/grain boundaries.
This (atomic-scale determined) reaction rate determines the rate of dislocation transmission/reflection and hence strongly influences the movement of dislocations plastic deformation of the material.
In extreme cases, where the reaction constant is effectively zero (i.e., the boundary is impenetrable),  dislocation motion is confined to within a grain.
Such a case is often assumed in  discrete dislocation dynamics (DDD) simulations of the flow stress~\citep{jiang2019effects}, where the flow stresses are significantly higher than the experimental observed. 
Typically, interfacial reactions lead to a relaxation of these flow stresses. 
A further example is that, lattice dislocation interactions with coherent twin boundaries accommodate dislocations from within the grains as well as dislocation transmission, leading to enhanced strength without compromising ductility~\citep{lu2004ultrahigh,li2010dislocation}. 
Hence, given different GBs with the distinct reaction constants, GB engineering can modify the mechanical response of polycrystals. 
\begin{figure*}[hbt]
\centering
\includegraphics[width=0.45\linewidth]{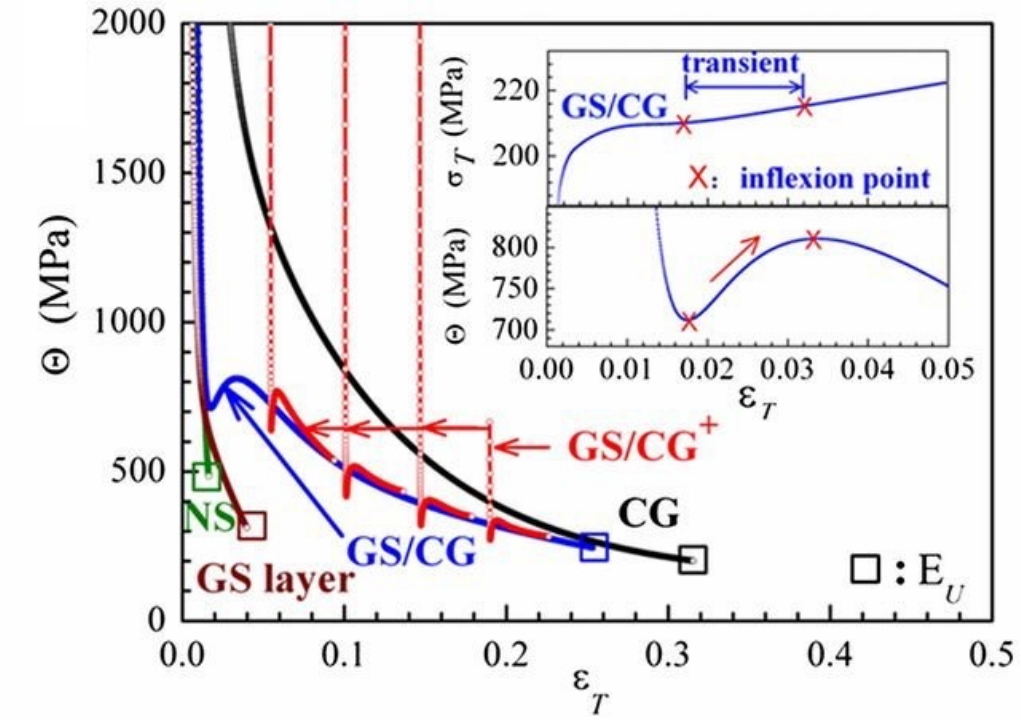}\hspace{-1.78em}%
\caption{\label{Theta_up_turn}
The $\Theta$ upturn phenomenon in gradient-structured coarse-grained interstitial free steel. 
CG: standalone homogeneous CG sample; GS layer: grain-size GS layer of 120-$\mu m$ thickness; GS–CG: sandwich sample of 1-$mm$ thickness. NS: freestanding, quasi-homogeneous nanostructured film of 20-$\mu m$ thickness peeled from the top surface of GS; GS–$\text{CG}^{+}$: the same sandwich sample subjected to unloading-reloading tensile testing at four separate strains of 0.05, 0.1, 0.15, and 0.2. 
All tensile samples were dog-bone-shaped, with a gauge dimension of $8 mm \times 2.5 mm$. 
Strain hardening rate ($\Theta \equiv d\sigma/d\epsilon$) vs. true strain ($\epsilon_{\text{T}}$) curves with permission from~\citep{wu2014extraordinary} (Copyright 2014 PNAS). 
The inset shows the transient response on the $\sigma_{\text{T}}$-$\epsilon_{\text{T}}$ curve of the GS–CG sample between two inflection points marked by ``$*$'' corresponding to the $\Theta$-upturn on its $\Theta$-$\epsilon_{\text{T}}$ curve. 
GS–$\text{CG}^{+}$ (red curve) also exhibits a $\Theta$-upturn upon each reloading.
}
\end{figure*}

Another intriguing phenomenon in polycrystals is how GBs and their distribution influence the strain hardening rate; in particular how  the  hardening rate varies with strain. 
The strain hardening rate defined as $\Theta \equiv \partial \sigma/ \partial \epsilon$, which is the slope of the stress-strain curve, quantifies the work hardening behavior, which is closely associated with  ductile fracture behavior. 
In tensile experiments, the strain hardening rate commonly decreases with increasing engineering strain. 
However, in gradient-structured coarse-grained (GS-CG) interstitial free (IF) steel, there is a transient hardening regime at small tensile strains, as depicted in the inset of Fig.~\ref{Theta_up_turn} \citep{wu2014extraordinary, wu2015heterogeneous} that leads to an up-turn in $\Theta$ with increasing strain (Fig.~\ref{Theta_up_turn}, inset). 
Unloading-reloading curves in the same material reveal a similar up-turn in $\Theta$ with each reloading for the GS-CG sample (red curves in Fig.~\ref{Theta_up_turn}). 
Interestingly, the GS-CG sample exhibits a slower reduction in $\Theta$ compared to the coarse grain (CG), but not gradient-structured, sample (Fig.~\ref{Theta_up_turn}). 
This indicates superior strain hardening retention in the GS-CG sample. 
In contrast, both a freestanding GS sheet and a CG sheet display a typical monotonic drop in $\Theta$ (see the GS layer and CG curves in Fig.~\ref{Theta_up_turn}), suggesting that this unique behavior arises only when these two layer types form an integrated  material.
One possible explanation for this up-turn in $\Theta$ is the scarcity of mobile dislocations at the onset of plastic deformation~\citep{wu2015heterogeneous}. 
Another explanation attributes the $\Theta$ up-turn to the high back stress resulting from the strain incompatibility caused by microstructural heterogeneity within the materials \citep{yang2021strain}. 
Although these qualitative explanations of the experimental observations provide some  insights, a clear, predictive model for this behavior remains elusive. 
We re-examine these observations using our mesoscale interface reaction model here to provide  more quantitative insights into these phenomena.

This paper employs numerical simulations to investigate the effects of interfaces on plasticity in polycrystalline materials. 
The main challenge lies in properly incorporating the microscopic interactions between interfaces and dislocations into the simulation model. 
We base the simulation model on the crystallographically-rigorous interface reaction model we proposed earlier (rather than on widely applied phenomenological approaches)
~\citep{yu2025mesoscale,yu2023application}.
We apply this mesoscale interface BC approach to a well-developed two-dimensional (2D) continuum dislocation dynamics (CDD) model
~\citep{leung2015new,ngan2017dislocation}. 
As our primary focus is on the mechanical effects of the interface BC, we consider a relatively simple 2D CDD model to describe the plasticity within the grains. 
We then apply this model to understand the influence of interfaces on strength and ductility.
We consider ductile fracture associated with stress localization  as well as brittle fracture resulting from dislocation pile-ups at the interface
~\citep{stroh1957theory, cottrell1965mechanics}. 
Our simulation results  demonstrate the nature of the competition between ductile and brittle fracture for different interfaces and loading conditions.

The remainder of this paper is organized as follows. 
Firstly, we briefly introduce the continuum dislocation dynamics model employed to describe the grain interiors and  review  our mesoscale interface boundary condition. 
We then present several  simulation cases for model validation.
Next, we apply the interface BC to study plastic deformation and mechanical properties for several GBs for a range of different interface reaction constants. 
We discuss the transition from ductile to brittle fracture under different conditions (strain rates, interface reaction constants).
Finally, we discuss the strain hardening rate versus strain as a function of interface reaction constants and dislocation generation rate to understand the $\Theta$ up-turn associated with GB strengthening described above.

\section{Methodology}\label{method}

\subsection{Definitions, Kinematics and the Interface Boundary Condition}\label{definition}
\begin{figure*}[hbt]
\centering
\includegraphics[width=0.6\linewidth]{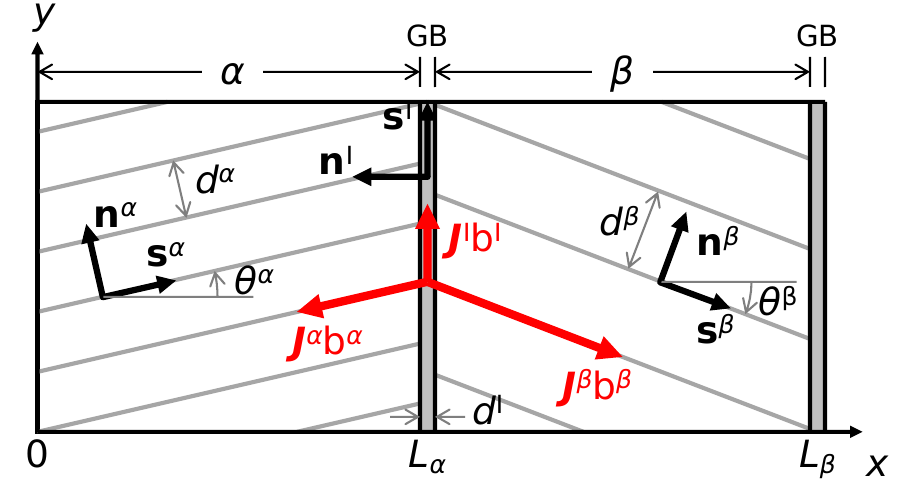}\hspace{-1.78em}%
\caption{\label{twodimension_model}
Configuration of a system composed by $\alpha$ and $\beta$ phases; the width of GB is $d^{\txI}$.
Periodic boundary condition is applied along the $x$- and $y$-axes. Gray lines denote the slip planes in each phase. Black vectors denote the slip direction and the normal of slip plane in each phase. 
Red vectors represent the Burgers vector fluxes from a point of an interface into $\alpha$ and $\beta$ phases and along the interface. 
If the fluxes are resulted from dislocation reaction, sum of the three red vectors should be zero.
}
\end{figure*}
Consider the periodic bicrystal model geometry  illustrated in Fig.~\ref{twodimension_model}.
We  focus on the single interface  at $x=0$ (labeled by ``I'') that is delimited by two phases (``$\alpha$'' and ``$\beta$''). 
There are multiple slip systems, characterized by a slip direction and a slip plane normal $(\mathbf{s}^{(i)}, \mathbf{n}^{(i)})$ ($(i)=\alpha_1, \alpha_2, \beta_1, \beta_2 ...$), within each phase.

The dislocation (not Burgers vector) density $\rho^{(i)}$ is the number of dislocations per unit  area (in the direction normal to the plane shown in Fig.~\ref{twodimension_model} on  slip system ``$(i)$''; the dislocation density vector is
\begin{equation}\label{define_dislocation}
\boldsymbol{\rho}^{(i)} = \rho^{(i)} \boldsymbol{\xi}, 
\end{equation}
where $\boldsymbol{\xi} \equiv \mathbf{s}^{(i)} \times \mathbf{n}^{(i)} $ is the dislocation line direction. 
The dislocation density vector evolution is simply (Maxwell-Faraday equation)~\citep{malygin1999dislocation,xia2015computational,ngan2017dislocation}
\begin{equation}\label{Faraday}
\dot{\boldsymbol{\rho}}^{(i)}
= - \nabla \times \left(\boldsymbol{\rho}^{(i)} \times \mathbf{v}^{(i)}\right), 
\end{equation}
where $\mathbf{v}^{(i)}$ is the dislocation velocity. 
Consider that the normal of the slip plane of a slip system is $\mathbf{n}^{(i)}$ and the dislocation line direction is always $\boldsymbol{\xi} = \mathbf{e}_z$. 
The positive slip direction is defined to be $\mathbf{s}^{(i)} = \mathbf{n}^{(i)} \times \mathbf{e}_z$; thus, $(\mathbf{s}^{(i)}, \mathbf{n}^{(i)}, \mathbf{e}_z)$ forms a local Cartesian coordinate system. 
Based on this coordinate system, $\boldsymbol{\rho}^{(i)} = \rho^{(i)}(x_s, x_n)\mathbf{e}_z$ and $\mathbf{v}^{(i)} = v^{(i)}(x_s, x_n)\mathbf{s}^{(i)}$, where $x_s$ and $x_n$ are  coordinates along the $\mathbf{s}^{(i)}$ and $\mathbf{n}^{(i)}$ axes and we assume that all quantities are uniform along  $\mathbf{e}_z$. 
In this case, 
\begin{equation}
\boldsymbol{\rho}^{(i)} \times \mathbf{v}^{(i)}
= \left|\begin{array}{ccc}
\mathbf{s}^{(i)} & \mathbf{n}^{(i)} & \mathbf{e}_z \\

v^{(i)} & 0 & 0
\end{array}\right|
= \rho^{(i)} v^{(i)} \mathbf{n}^{(i)}. 
\end{equation}
\begin{equation}
\nabla \times \left(\boldsymbol{\rho}^{(i)} \times \mathbf{v}^{(i)}\right)
= \nabla \times \left(\rho^{(i)} v^{(i)} \mathbf{n}^{(i)}\right)
= \frac{\partial(\rho^{(i)} v^{(i)})}{\partial x_s} \mathbf{e}_z
= \left(\cos\theta^{(i)}\frac{\partial(\rho^{(i)} v^{(i)})}{\partial x} + \sin\theta^{(i)}\frac{\partial(\rho^{(i)} v^{(i)})}{\partial y}\right) \mathbf{e}_z,
\end{equation}
where $\theta^{(i)}$ is the angle between $\mathbf{s}^{(i)}$ and $\mathbf{e}_x$, such that Eq.~\eqref{Faraday} becomes
\begin{equation}\label{divflux}
\dot{\rho}^{(i)}
= -\frac{\partial(\rho^{(i)} v^{(i)})}{\partial x_s}
= -\cos\theta\frac{\partial(\rho^{(i)} v^{(i)})}{\partial x}
- \sin\theta\frac{\partial(\rho^{(i)} v^{(i)})}{\partial y}.
\end{equation}
This equation describes conservation of dislocations on each slip system. 
Note that the dislocation density and velocity can be expressed in either the slip system $(\mathbf{s}^{(i)},\mathbf{n}^{(i)},\mathbf{e}_z)$ or Cartesian  $(\mathbf{e}_x,\mathbf{e}_y,\mathbf{e}_z)$ coordinate systems.
To describe dislocation generation/annihilation we distinguish between dislocations of opposite sign. 
A ``$+$/$-$'' dislocation on a particular slip system moves in the positive/negative slip direction when the resolved shear stress $\tau^{(i)} > 0$ (depending on the definition of $\mathbf{n}^{(i)}$). 
The dynamics of ``$+$'/$-$'' dislocation densities on a particular slip system are described by 
$\dot{\rho}^{(i)}_+ = -{\partial(\rho^{(i)}_+ v^{(i)}_+)}/{\partial x_s}$
and
$\dot{\rho}^{(i)}_- = -{\partial(\rho^{(i)}_- v^{(i)}_-)}/{\partial x_s}$.
The resolved shear stress (RSS) $\tau^{(i)}$ drives slip on each slip system~\citep{johnston1959dislocation,ngan2017dislocation,dickel2014dipole}; assuming a power law force-velocity relation and a Peierls/friction stress $\tau_{\text{f}}$, the dislocation velocity is 
\begin{equation}\label{TW_kineticslaw_modified}
v^{(i)}_+ =-v^{(i)}_-=
    \begin{cases}
       v_0\usgn(\tau^{(i)} - \tau_{\text{f}}) \left|\frac{\tau^{(i)}-\tau_{\text{f}}}{\tau_0}\right|^n, & \tau^{(i)} - \tau_f > 0 \\
      0, & \tau^{(i)} - \tau_f \le 0,\\
    \end{cases}
\end{equation}
where $v_0$ and $\tau_0$ are constants, $\tau_{\text{f}}$ represents the Peierls stress, which is the minimum shear stress required to move a single dislocation of unit length in a perfect crystal.
The annihilation of two opposite sign dislocation can also be expressed as
\begin{equation}\label{TW_dotrhoann}
\dot{\rho}_+^{(i),\text{ann}}
= \dot{\rho}_-^{(i),\text{ann}}
= - \rho^{(i)}_+ \rho^{(i)}_- r_\txc \left|v^{(i)}_+ - v^{(i)}_-\right|, 
\end{equation}
where $v^{(i)}_{+/-}$ denote the velocity of positive and negative dislocations respectively and $r_\txc$ is the capture radius which approximately equals to the dislocation core size.  
The generation of pairs of dislocation can be written as~\citep{kocks1975progress} 
\begin{equation}\label{dotrhogen_two}
\dot{\rho}^{(i),\text{gen}}_+
= \dot{\rho}^{(i),\text{gen}}_-
=\eta |\tau^{(i)}|^{m}, 
\end{equation}
where $m$ and $\eta$ are the  source term exponent and source intensity respectively and can be treated as parameters.
Thus, the evolution of dislocation density is
\begin{equation}\label{evolution_of_dd}
    \dot{\rho}^{(i)}_{+/-}=-\frac{\partial(\rho^{(i)}_{+/-}v^{(i)}_{+/-})}{\partial x_s}+\dot{\rho}^{(i),\text{gen}}_{+/-}+\dot{\rho}_{+/-}^{(i),\text{ann}}.
\end{equation}
All  of the quantities  in this equation are functions of space and time $\{x,y,t\}$.

Most dislocations gliding to the interface from the adjoining grain are blocked by and pile up against the interface.
However, some  may react at the interface.
In order to quantitatively describe the  dislocation reaction kinetics, we proposed a mesoscale interface BC  based up the maximum rate of entropy production principle~\citep{yu2025mesoscale}.
Consider the general case in which there are $n$ ($n \ge 4$) non-colinear slip systems in the two grains; the interface can also be treated as an additional  slip system. 
Arbitrary reaction ``($k$)''  may occur among any four of the $n$ slip systems.
The kinetics of reaction ``($k$)'' may be described by
\begin{equation}\label{dissipation_multi}
\mathbf{J}_{(k),\txI}
=
\kappa_{(k)} \mathbf{B}^{-1} (\mathbf{c}\otimes\mathbf{c})_{(k)} \mathbf{B} \boldsymbol{\tau}_{(k),\txI} 
\boldsymbol{\rho}_{(k),\txI},
\end{equation}
where the subscript ``$(k)$'' and ``$\txI$'' represent  reaction ``$(k)$'' and the interface where all quantities are evaluated at the interface, $\kappa_{(k)}$ is the reaction constant for reaction ``$(k)$''.
$\mathbf{J}_{(k),\txI} \equiv \left(J^{(1)}, J^{(2)}, J^{(3)}, J^{(4)}\right)^\text{T}$ and $\boldsymbol{\rho}_{(k),\txI} \equiv \left(\rho^{(1)}, \rho^{(2)}, \rho^{(3)}, \rho^{(4)}\right)^\text{T}$ are the generalized dislocation flux and density. 
 $\mathbf{c} \equiv \left(c^{(234)}, c^{(314)}, c^{(124)}, c^{(132)} \right)^\text{T}$ only depends on the crystallographic orientation of the two phases meeting at the interface, where $c^{(ijk)} \equiv \mathbf{s}^{(i)} \cdot (\mathbf{s}^{(j)} \times \mathbf{s}^{(k)})$ ($i,j,k=1,2,3,4$).
 $\mathbf{B} \equiv \mathrm{diag}\left(b^{(1)}, b^{(2)}, b^{(3)}, b^{(4)}\right)$ is the list of Burgers vectors and $\boldsymbol{\tau}_{(k),\txI} \equiv \mathrm{diag}\left(\tau^{(1)}, \tau^{(2)}, \tau^{(3)}, \tau^{(4)}\right)$ represents the shear stresses resolved on the individual slip systems (RSS). 
The fluxes described by Eq.~\eqref{dissipation_multi} only account for  dislocation reaction ``$(k)$''.
Similar relations apply to all combinations of four slip systems. 
The total Burgers vector flux, due to all $C_{n}^{4}\equiv m$ reactions, is
\begin{equation}\label{rbc10}
\bar{\mathbf{J}}_{\txI}
= \sum_{n=1}^{m} \bar{\mathbf{J}}_{(n),\txI}
= \bar{\mathbf{B}}^{-1}
\left(\sum_{n=1}^{m} \kappa_{(n)} \overline{(\mathbf{c}\otimes\mathbf{c})}_{(n)}\right)
\bar{\mathbf{B}}\bar{\mathbf{T}}_{\txI}\bar{\boldsymbol{\rho}}_{\txI}.
\end{equation}
\noindent where 
\begin{align}
 \bar{\mathbf{B}} &\equiv \mathrm{diag}(b^{(1)},b^{(2)}, \cdots,b^{(n)}), \quad
 \bar{\mathbf{T}}_\txI \equiv \mathrm{diag}(\tau^{(1)}, \tau^{(2)},\cdots,\tau^{(n)}),  \nonumber  \\
\bar{\boldsymbol{\rho}}_\txI &\equiv (\rho^{(1)}, \rho^{(2)},\cdots ,\rho^{(n)})^\text{T}, \quad 
\bar{\mathbf{J}}_\txI \equiv (J^{(1)}, J^{(2)}, \cdots ,J^{(n)})^\text{T} \nonumber .
\end{align}
This general form Eq.~\eqref{rbc10} of the interface BC can be applied to all dislocation reactions involving any number of slip systems. 

In the simulations  below, we apply periodic boundary conditions in both  $x$- and $y$-direction (i.e.,  dislocations exiting one side of the simulation  cell re-enter  on the opposite side).
In the later analysis, we express physical quantities based in dimensionless form:
\begin{equation}\label{defdimensionless}
\tilde{\rho} \equiv \rho L_x^2, \quad
\tilde{x} \equiv x/L_x, \quad
\tilde{t} \equiv v_0 t/ L_x, \quad
\tilde{v} \equiv v / v_0, \quad
\tilde{\tau} \equiv \tau / K, 
\end{equation}
where $K = \mu / [2 \pi (1 - \nu)]$, $L_x$ is the length of the simulation cell in the $x$-direction.
For simplicity, we  omit the tilde in the dimensionless quantities below.
The  numerical simulation details and parameters are presented in~\ref{numerical_simulations} and in  Table~\ref{table:1} respectively.
It should be noted, in the Table~\ref{table:1}, the superscript $i$ denoting the slip system is omitted since, for these  simulations, we assume  the parameters are identical for all slip systems.

\subsection{Validation: impenetrable interfaces}\label{Neumann}

\begin{figure}[htb]
\centering
\includegraphics[width=1.0\linewidth]{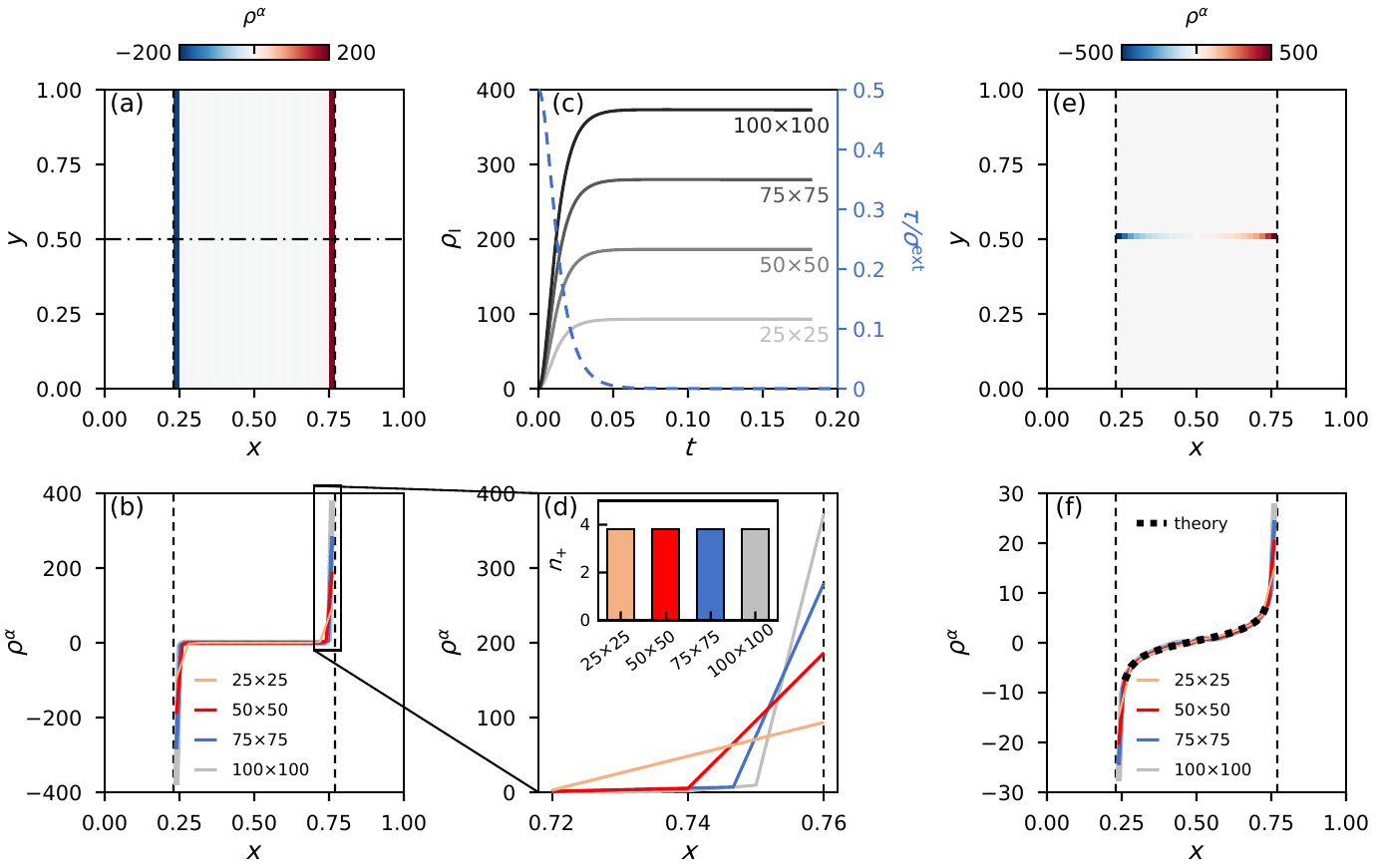}\hspace{-1.78em}
 \caption{\label{td_gauss}
Simulation results for a periodic bicrystal geometry under an applied stress of $\sigma^\text{ext}=0.01$.
(a) shows the distribution of dislocation density over the whole region at final states for simulations with a  $50 \times 50$ mesh;
(b) is the distribution of dislocation density along the $x$-direction with different mesh  sizes;
(c) shows the evolution of dislocation density at the interface $\rho_\txI$ (i.e., the  dislocation density at the GB, $x=0.76$) and averaged resolved shear stress $\tau$ within the grain for different meshes; 
(d) enlargement of the dislocation density spatial distribution from (b) near the interface ($x=0.76$) and the inset shows the magnitude of the dislocation density localized within one mesh point of the interface; 
(e) the distribution of dislocation density  for the case where there is a single dislocation source located at the centre of the unit cell (mesh size $50 \times 50$); 
(f) shows the distribution of dislocation density along $x$-axis for the case shown in (e) for several mesh sizes.
}
\end{figure}

Consider first the case in which dislocations cannot cross or react with the interface; this corresponds to a  Neumann boundary condition (BC) at the interface (i.e., shown as the vertical  dashed lines in the periodic simulation cell in Fig.~\ref{td_gauss}(a)). 
The dislocation sources are assumed to exist at every (mesh) point in the microstructure; here, we employ a $50 \times 50$ mesh.
When an external force is applied, the resolved shear stress (RSS) activates the dislocation sources and drives the glide of opposite signed dislocations in opposite directions. 
Since there is no dislocation flux crossing the boundary, all dislocations pile up against the interface, as seen in Fig.~\ref{td_gauss}(a, b). 
Qualitatively, the distribution of dislocation density along the $x$-axis exhibits a sharp increase close to the GB and decays rapidly with distance from the GB, as depicted in Fig.~\ref{td_gauss}(b)~\citep{scardia2014mechanics}. 
Figure~\ref{td_gauss}(c) shows that the total RSS $\tau$, denoted by the dashed blue line, decays to zero over time. 
As the dislocations pile up at the interface, they  induce a back stress that counteracts the external RSS, shutting off the dislocation sources.


From Fig.~\ref{td_gauss}(b), we observe that the maximum dislocation density in the pile-up near the grain boundary increases with decreasing mesh size (inverse of the number of mesh points in each direction).
This is simply because the dislocation density decays over a single mesh spacing - as seen in Fig.~\ref{td_gauss}(c); the total dislocation density within a single mesh spacing of the interface is constant and the maximum dislocation density is inversely related to the mesh spacing.
In other words,  the total number of positive dislocations within the right half region denoted by $S_\text{r}$, defined as $n_{+}=\sum_{S_\text{r}}\rho^{\alpha} \Delta x \Delta y$ (where $\Delta x$ and $\Delta y$ are the mesh spacings), remains constant as the inset shows in Fig.~\ref{td_gauss}(d) when the mesh size is varied.

Now, consider the case where the simulation cell contains only one point source located at the center $(x',y')$. 
The source term may be described as $\dot{\rho}^\text{gen}_+ = \dot{\rho}^\text{gen}_- = \eta |\tau|^{m} \delta (x-x', y-y')$, where $\delta (x-x', y-y')$ represents the Dirac delta function. 
The unit cell is also periodic in both the $x$- and $y$- directions with Neumann boundary conditions. 
The slip system in the center crystal is $\theta^\alpha = 0^\circ$; all other parameters are the same as shown in Table \ref{table:1}.
As shown in Figure \ref{td_gauss}(e), positive and negative dislocations glide along the slip plane and pile up against the interface. 
Since there is only one slip plane in the unit cell, the distance between sources and planes along which dislocation glides is large; i.e., set by the periodic unit cell size in the $y$-direction, $100b$ (where $b$ is the magnitude of Burgers vector). 
The dislocation density profile obtained from our simulation results can be compared to the theoretical result derived based on a single slip plane~\citep{hirth1982theory}.
As seen in Fig.~\ref{td_gauss}(f), the agreement between the simulation and the theory is excellent.
This validates the basic simulation model.

We now examine  the effect of interfaces on the mechanical properties of materials under uniaxial tension (in the $y$-direction; i.e., the model is loaded at a constant nominal strain rate $\dot{\epsilon}$. 
The model undergoes both elastic and plastic deformation. 
The plastic strain rate associated with slip system $i$ can be calculated using the Orowan equation as $\dot{\gamma}^{(i)}= (\rho^{(i)}_{+} + \rho^{(i)}_{-})v^{(i)}_{+}b^{(i)}$, where $\rho_{+/-}^{(i)}$ and $v_{+/-}^{(i)}$, denote  the  density and velocity of positive/negative dislocations \citep{cai2018imperfections}. 
The total plastic strain rate tensor can be obtained by summing over all slip systems and expressed as \citep{cleveringa1997comparison}:

\begin{equation}\label{plastical_strain_rate}
    \dot{\boldsymbol{\epsilon}}^{\text{p}} = \sum_{i}\frac{\dot{\gamma}^{(i)}}{2}(\mathbf{s}^{(i)} \otimes \mathbf{n}^{(i)} + \mathbf{n}^{(i)} \otimes \mathbf{s}^{(i)}),
\end{equation}
where the summation is taken over all slip systems. 
The external loading stress rate is
$\dot{\sigma} = E\dot{\epsilon}^{\text{e}}= E (\dot{\epsilon} - \dot{\epsilon}^{\text{p}})$,
where $\dot{\epsilon}$ is the nominal strain rate, $\dot{\epsilon}^{\text{p}}$ is the corresponding component of the plastic strain rate tensor $\dot{\boldsymbol{\epsilon}}^{\text{p}}$ along the loading direction, and $E$ is the Young's modulus of the material \citep{el2009role,zhang2021dislocation}. 

\subsection{Fracture Criteria} 


The  material may undergo  ductile or brittle fracture on loading process \citep{zhu2010ultra}. 
We consider ductile failure in tensile loading in terms of the  necking criterion proposed by Hart  \citep{hart1967theory}. 
In this approach, strain localization occurs when the work hardening in the material is unable to compensate for  the increment of loading:
\begin{equation}\label{Hart_criterion}
\Theta \geq (1-m)\sigma,
\end{equation}
where $\sigma(\epsilon, \dot{\epsilon})$ is the stress as a function of strain $\epsilon$ and strain rate $\dot{\epsilon}$,   $\Theta \equiv (\partial \sigma / \partial \epsilon)_{\dot{\epsilon}}$ is the strain hardening rate, and $m \equiv (\partial \ln \sigma / \partial \ln \dot{\epsilon})_{\epsilon}$ is the strain rate sensitivity. 
While the Hart criterion is applicable to rate-dependent materials, for rate-independent materials, the Consid\`ere condition may be applied
$\Theta \geq \sigma /(1 + \epsilon)$,
where the stress $\sigma(\epsilon)$ is rate-independent and only dependent on strain $\epsilon$. 
We focus here on the Hart criterion.

\begin{figure*}[hbt]
\centering
\includegraphics[width=0.85\linewidth]{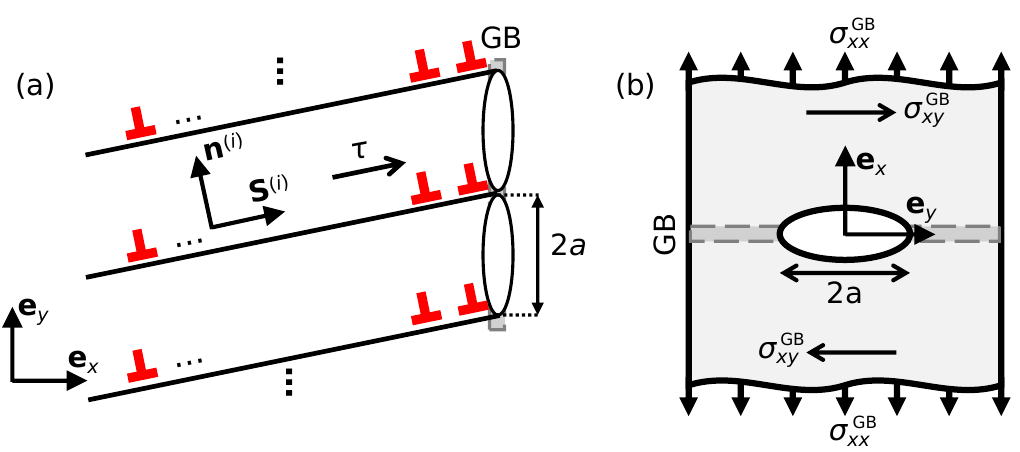}\hspace{-1em}%
\caption{\label{brittle_GB_model}
The fracture nucleation scheme based on the linear elastic fracture model with infinite slip plane along $y$-direction. 
(a) There are $n$ piled up dislocations on each slip plane under resoled shear stress $\tau$, leading to microcrack initiation near the interface/GB for two-dimensional model; 
(b) the stress field induced by piled up dislocation acting at the GB with a microcrack of length $2a$.
}
\end{figure*}

For brittle fracture, on the other hand, microplasticity is limited and cracks nucleated at interface coalesce to lead to sudden fracture.
In the brittle fracture dislocation theory introduced by Stroh, fracture occurs when the stress at the end of a dislocation pile-up at the interface exceeds the stress required to decohere the interface \citep{stroh1957theory,liu2023microstructure}. 
While there are other dislocation models for fracture  \citep{cottrell1965mechanics,smith1967crack}, these criteria cannot be applied directly to our model because they focus on single slip planes. 
Our continuum model considers an infinite array of possible slip planes as Fig.~\ref{twodimension_model} shows.

As illustrated in Figs.~\ref{brittle_GB_model}(a) and (b), we consider a microcrack of length $2a$ produced by the piled up dislocations at the GB, and analyze the crack propagation under the shear and normal stresses $\sigma_{xx}^{\text{GB}}$, $\sigma_{xy}^{\text{GB}}$ and $\sigma_{yz}^{\text{GB}}$ associated with the dislocations from within the grains (and the GB disconnections).
In the linear elastic theory of fracture, the crack will extend when the energy release rate is sufficient to overcome the change in surface/interface energy, which can be expressed as~\citep{weertman1996dislocation}
\begin{equation}\label{criterion}
    \frac{K_\txI^2(1-\nu)}{2G}+\frac{K_{\txI\txI}^2(1-\nu)}{2G}+\frac{K_{\txI\txI\txI}^2}{2G}=2\gamma_{\text{eff}}.
\end{equation}
where $K_\txI$, $K_{\txI\txI}$ and $K_{\txI\txI\txI}$ are the mode I, II and III stress intensity factors associated with the total stress, respectively.
$G$ and $\nu$ are the shear modulus and Possion ratio, respectively.
$\gamma_{\text{eff}}=2\gamma-\gamma_{\txI}$ represents the effective work for GB decohesion, $\gamma$ and $\gamma_\txI$ are the fracture surface energy of the materials and surface energy of GB.

In this study, the mode III stress intensify factor $K_{\txI\txI\txI}$ is not relevant because of the two-dimensional nature of the model.
The stress intensity factors for an assumed straight crack of length $2a$ can be determined from the stress  at the GB $\sigma_{xx}^{\text{GB}}$ and $\sigma_{xy}^{\text{GB}}$:
\begin{equation}\label{SIF}
    K_{\txI}=\sigma_{xx}^{\text{GB}} \sqrt{\pi a}, \quad K_{\txI\txI} = \sigma_{xy}^{\text{GB}} \sqrt{\pi a}, \quad K_{\txI\txI\txI}=0,
\end{equation}
where the substitle ``GB'' represents the stress evaluated at the GB.
When there are $n$ dislocations piled up at the GB under the resolved shear stress $\tau$, a microcrack with length $a$ will grow to release the local elastic energy.
In this continuum model, the number of piled up dislocation $n$ can be calculated by the dislocation density times pile-up length along slip direction.
It is worth mentioning that the size of microcrack is quite small compared with the crystal dimensions, and spontaneous growth of microcrack will not occur unless the material is under the action of applied stress.
The crack length associated with dislocation pile-up can be written as~\citep{sarfarazi1987microfracture}
\begin{equation}\label{microcrack}
    2a=\lambda (n^{\text{GB}}_x)^2,
\end{equation}
where $\lambda$ is a constant related to the material properties, $n^{\text{GB}}_x$ represents the number of piled up dislocations at the GB with Burgers vector $\mathbf{b}$ projected in the $x$-direction.
The piled up dislocation density $n^{(i)}$ with Burgers vector $\mathbf{b}^{(i)}$ can be projected onto the $x$- and $y$-directions by $n^{(i)}_x b^{(i)} \equiv n^{(i)} \mathbf{b}^{(i)} \cdot \mathbf{e}_x= n^{(i)} \cos \theta^{(i)}b^{(i)}$ and $n^{(i)}_y b^{(i)} \equiv n^{(i)} \mathbf{b}^{(i)} \cdot \mathbf{e}_y = n^{(i)} \sin \theta^{(i)} b^{(i)}$, where $\theta^{(i)}$ is the angle between the slip direction $\mathbf{s}^{(i)}$ and the $x$-axis and $b^{(i)}$ is the magnitude of the Burgers vector $\mathbf{b}^{(i)}$.
Only the $x$-component contributes to microcrack propagation; the $y$-component of the Burgers vectors contributes to GB sliding.
The $x$-component of the Burgers vector of total dislocation near the GB is $n^{\text{GB}}_x=n_x^\alpha+n_x^\beta$, meaning that piled up dislocations from both sides of the GB/interface can contribute to initiate a microcrack.

Substituting Eq.~\eqref{SIF} and Eq.~\eqref{microcrack} into Eq.~\eqref{criterion}, yields
\begin{equation}\label{brittle_criterion}
    M \equiv \left[(\sigma_{xx}^{\text{GB}})^2 + (\sigma_{xy}^{\text{GB}})^2\right] (n^{\text{GB}}_x)^2 = \frac{8\gamma_{\text{eff}}}{\pi \lambda (1-\nu)},
\end{equation}
where the left-hand side can be expressed in terms of the stress concentration and the density of dislocations at the GB, and we can define $M \equiv \left[(\sigma_{xx}^{\text{GB}})^2+ (\sigma_{xy}^{\text{GB}})^2\right](n^{\text{GB}}_x)^2$ for simplicity.
The right-hand side is only related to the properties of the materials and can be considered as strength of GB, and is assumed to be a constant $M^* \equiv 8\gamma_{\text{eff}}/\pi \lambda (1-\nu)$ in the later analysis.
During our simulation, the stress components at the GB, $\sigma_{xx}^{\text{GB}}$ and $\sigma_{xy}^{\text{GB}}$, and the local dislocation density at the GB $n^{\text{GB}}_x$ can be obtained at each step during the simulation process.

In summary, we propose a microcrack propagation criterion for our two-dimensional model with infinite slip planes based on linear elastic theory of fracture. 
The critical condition for crack propagation at the GB is given by Eq.~\eqref{brittle_criterion}, which relates the stress components $\sigma_{xx}^{\text{GB}}$ and $\sigma_{xy}^{\text{GB}}$ to the surface/interface energies, material properties, and the local dislocation density. 
This criterion allows us to predict  crack propagation in our model for the investigation of  brittle fracture.
It should be noted that we monitor both failure criteria, associated with necking and crack nucleation at the GB, during the simulations.
Later, we will study which one dominates and determine which mechanism governs the failure of the material during the tensile test.

\section{Results and discussions}\label{results}
Interfaces can affect both the plastic and fracture response of metallic alloys \citep{zhu2023heterostructured, huang2018interface, zhu2010ultra}. 
Herein, we examine the effects of interfaces on plastic deformation and fracture using our proposed interface BC and crystal plasticity model.

\subsection{Effects of reaction constant on mechanical properties}
\begin{figure}[hbt]
\centering
\includegraphics[width=1.0\linewidth]{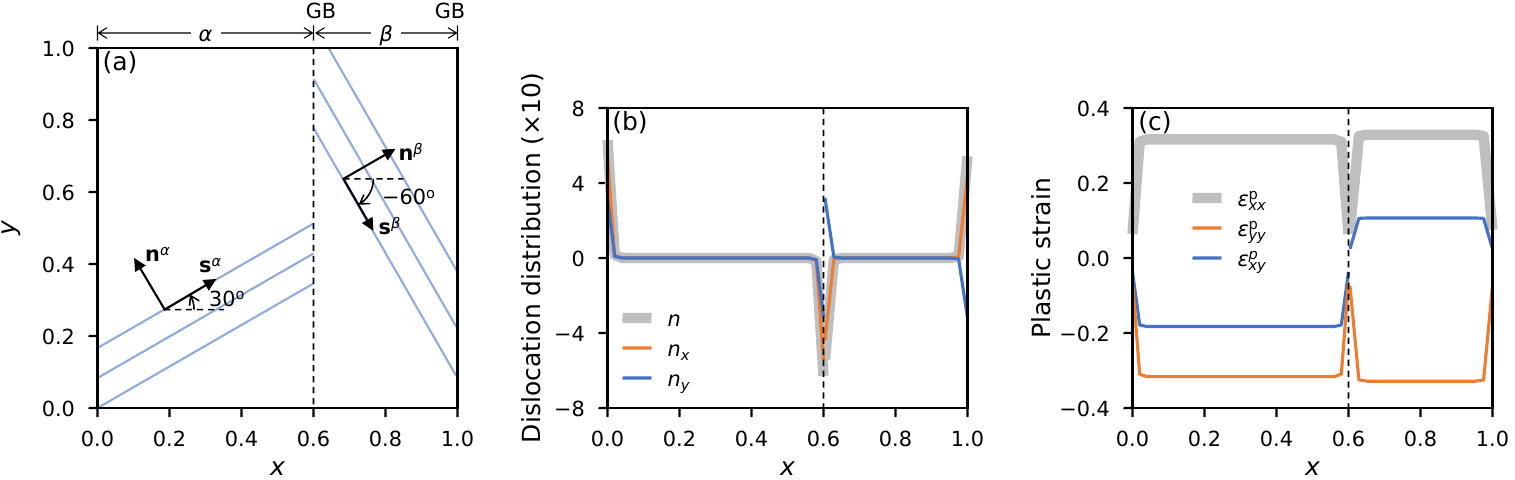}\hspace{-1em}%
\caption{\label{robin_config}
(a) shows that the schematic of the model, consisting of $\alpha$ and $\beta$ phase which has only slip system respectively.
For the case with reaction constant $\kappa=10$, (b) shows the dislocation distribution along $x$-axis and the vertical black dashed lines represent the edges of GB; 
The gray, yellow and blue lines represent dislocation density, dislocation density projected into $x$- and $y$-axes respectively. 
(c) shows the distribution of three components of plastic strains. 
}
\end{figure}

We explore the influence of the reaction constant, $\kappa$, in a two-phase system. 
Each phase contains only one slip system with misorientation angles of $\theta^{\alpha} = 30^\circ$ and $\theta^{\beta} = -60^\circ$, respectively. 
Therefore, the interface BC is a Robin boundary condition (reaction BC) with varying reaction constants, $\kappa$. 
The dislocation sources are assumed at each site, and an applied force is exerted along the $x$-axis, which is perpendicular to the interface, as shown in Fig.~\ref{robin_config}(a).

From Fig.~\ref{robin_config}(b), the distribution of dislocations along the $x$-axis is shown, as it is uniform in the $y$-direction. 
When an external force is applied to the model, dislocations with different signs move in opposite directions and are blocked by the GB denoted by the black dashed line in Fig.~\ref{robin_config}(a). 
The accumulation of dislocations at the interface induces back stress within the grain, resulting in strain hardening of the material. 
The dislocation number $n^{(i)}$ is measured under the local coordinate system of each slip system.
To measure the dislocation number in the global coordinate system, $x$ and $y$ component of total dislocations $n_x$ and $n_y$ are shown in Fig.~\ref{robin_config}(b). 
The corresponding plastic strains are also plotted in Fig.~\ref{robin_config}(c) respectively.
\begin{figure}[hbt]
\centering
\includegraphics[width=0.98\linewidth]{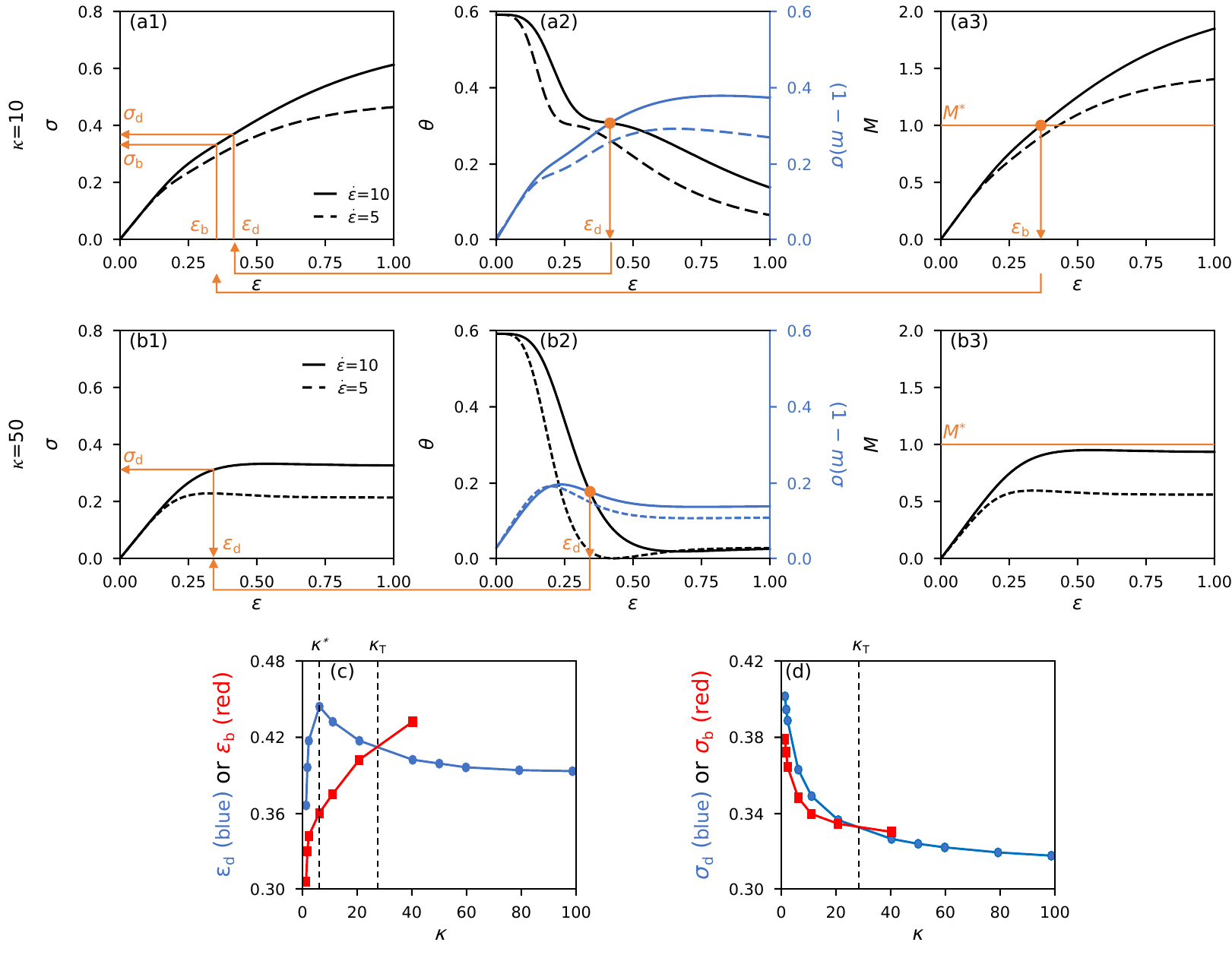}\hspace{-1em}%
\caption{\label{stress_strain_curve}
Comparison of ductile and brittle fracture strains predicted by Hart's and Weertman's criteria, respectively.
The first two row correspond to the results under the reaction constant $\kappa=10$ and $\kappa=50$.
(a1) and (b1) Stress-strain curve of material under different strain rate $\dot{\epsilon}$ denoted by the solid and dashed lines respectively. 
(a2) and (b2) The black and blue lines represent strain hardening $\Theta$ and $(1-m)\sigma$ respectively.
(a3) and (b3) Evolution of the local value  $M \equiv \left[(\sigma_{xx}^{\text{GB}})^2+ (\sigma_{xy}^{\text{GB}})^2\right](n^{\text{GB}}_x)^2$ near the GB under different strain rate $\dot{\epsilon}$ and threshold value for interfacial strength denoted by orange horizontal line. 
(c) The ductile strain $\epsilon_{\text{d}}$ and brittle strain $\epsilon_{\text{b}}$ are plotted under different reaction constants $\kappa$ respectively.
(d) The strength of the material corresponds to the brittle and ductile fracture under varying reaction constants.
}
\end{figure}

In Fig.~\ref{stress_strain_curve}, the first two rows correspond to the results with reaction constants $\kappa=10$ and $\kappa=50$, respectively. 
Stress-strain curves are plotted under different strain rates ($\dot{\epsilon}=5,10$), denoted by solid and dashed lines in Fig.~\ref{stress_strain_curve}(a1). 
These curves show that the strength increases with the applied strain rate.
The onset strain of necking is shown in Fig.~\ref{stress_strain_curve}(a2) and (b2). 
The black and blue lines represent the strain hardening rate $\Theta$ (left-hand side of the Hart criterion) and $(1-m)\sigma$ (right-hand side of the Hart criterion) in Eq.~\eqref{Hart_criterion}, respectively. 
The intersection of these lines represents the onset point of necking. 
The necking process can be interpreted as follows: in the first stage, the strain hardening of the material grows more rapidly than stress localization, allowing the material to remain stable. 
As the external load continues to increase and exceeds the intersection point, the material starts to collapse because the rate of stress localization becomes larger than the strain hardening rate. 
Considering crack nucleation criterion, Fig.~\ref{stress_strain_curve}(a3) and (b3) depict the evolution of local value $M \equiv \left[(\sigma_{xx}^{\text{GB}})^2+ (\sigma_{xy}^{\text{GB}})^2\right](n^{\text{GB}}_x)^2$ near the GB. 
The crack nucleates once the local value $M$ reaches a certain threshold value $M^*$ that equals R.H.S. in Eq.~\eqref{brittle_criterion} (orange horizontal line in Fig.~\ref{stress_strain_curve}(a3) and (b3) which sets the critical $M^*$ value to one in this illustrative example). 
Once the ductile and brittle fracture strains are determined, the corresponding strength of the material can be obtained through the stress-strain curves, as shown by the orange lines in Fig.~\ref{stress_strain_curve}(a1) and (b1).

The dislocation fluxes represent the rate of dislocations reacted per unit length per unit time at the interface. 
The reaction constant $\kappa$, as shown in Eq.~\eqref{dissipation_multi}, controls the magnitude of the dislocation fluxes reacting at the interface.
Therefore, it is crucial to explore the effect of the reaction constant $\kappa$ on the mechanical properties of the materials. 
Similar to the previous cases, stress-strain curves can be obtained for different reaction constants $\kappa$. 
Additionally, the onset strain of necking and crack nucleation can be determined by considering corresponding failure criteria.

On the one hand, there is a peak value $\kappa^*$ for the ductility of the material represented by the blue line with the increase of the reaction constant $\kappa$ as shown in Fig.~\ref{stress_strain_curve}(c). 
When the reaction constant $\kappa$ increases until it exceeds $\kappa^*$, more dislocations can transmit across the interface, resulting in larger plastic deformation and increased ductility of the material. 
However, when $\kappa$ continues to increase and exceeds $\kappa^*$, the reaction dislocation flux becomes too large, leading to minimal piled up dislocations at the interface. 
As a result, the strain hardening term $\Theta$ in the Hart criterion Eq.~\eqref{Hart_criterion} becomes significantly smaller, which in turn decreases the ductility of the material. 
It should be noted that our model does not consider any other strengthening mechanisms such as dislocation hardening, solid solution strengthening, precipitation hardening, and dispersion hardening, apart from the GB strengthening. 
In the extreme case where the reaction constant approaches infinity ($\kappa \rightarrow \infty$), the interface can hardly block any dislocations, resembling a single crystal. In such a case, no strain hardening occurs, and necking quickly initiates once an external force is applied.

On the other hand, as the dislocations pile up against the interface, the increased local value $M$ near the interface which is function of piled up dislocation and local stress may lead to crack nucleation at the interface as criterion shown in Eq.~\eqref{brittle_criterion}. 
Once the local value $M$ exceeds a certain threshold value ($M \ge M^*$), the material undergoes brittle fracture at the interface, and the corresponding strain $\epsilon_{\mathrm{b}}$ can be considered as the onset strain of crack nucleation, as shown in Fig.~\ref{stress_strain_curve}(a3) and (b3).
Comparing the two strains mentioned above, if $\epsilon_{\mathrm{d}} < \epsilon_{\mathrm{b}}$, it can be classified as ductile fracture; otherwise, it is brittle fracture ($\epsilon_{\mathrm{d}} > \epsilon_{\mathrm{b}}$). 
With an increase in the reaction constant $\kappa$, there is a transition point $\kappa_{\text{T}}$ for the failure strain, as depicted in Fig.~\ref{stress_strain_curve}(c). 
Below this transition point, the fracture mode is dominated by brittle fracture. 
When $\kappa$ gradually increases but remains below the transition point ($\kappa \le \kappa_{\text{T}}$), more piled up dislocations can be transmitted to another phase, releasing the local tensile stress near the interface and resulting in an increase in the failure strain $\epsilon_{\mathrm{b}}$ governed by brittle fracture. 
When the reaction constant $\kappa$ continues to increase and surpasses the critical point $\kappa_{\text{T}}$, the fracture mechanism transitions from brittle to ductile modes, causing a gradual decrease in the failure strain. 
It is worth mentioning that when $\kappa$ is sufficiently large, brittle fracture may vanish, as shown in Fig.~\ref{stress_strain_curve}(b3), since the local value $M$ near the interface induced by minimal piled up dislocations cannot exceed the interfacial strength $M^*$.

From Fig.~\ref{stress_strain_curve}(d), the ultimate strength of the material also exhibits a transition point $\kappa_{\text{T}}$ corresponding to the failure strain. 
Initially, the strength is controlled by brittle fracture and then transitions to ductile fracture. 
As the reaction constant $\kappa$ increases, more dislocations react at the interface, resulting in fewer piled up dislocations. 
The reduced number of piled up dislocations leads to less strengthening of the material, causing the strength to decrease with an increase in the reaction constant $\kappa$, as shown in Fig.~\ref{stress_strain_curve}(d).

\subsection{Transition of brittle and ductile fracture}\label{discussion_bdt}
\begin{figure}[hbt]
\centering
\includegraphics[width=0.90\linewidth]{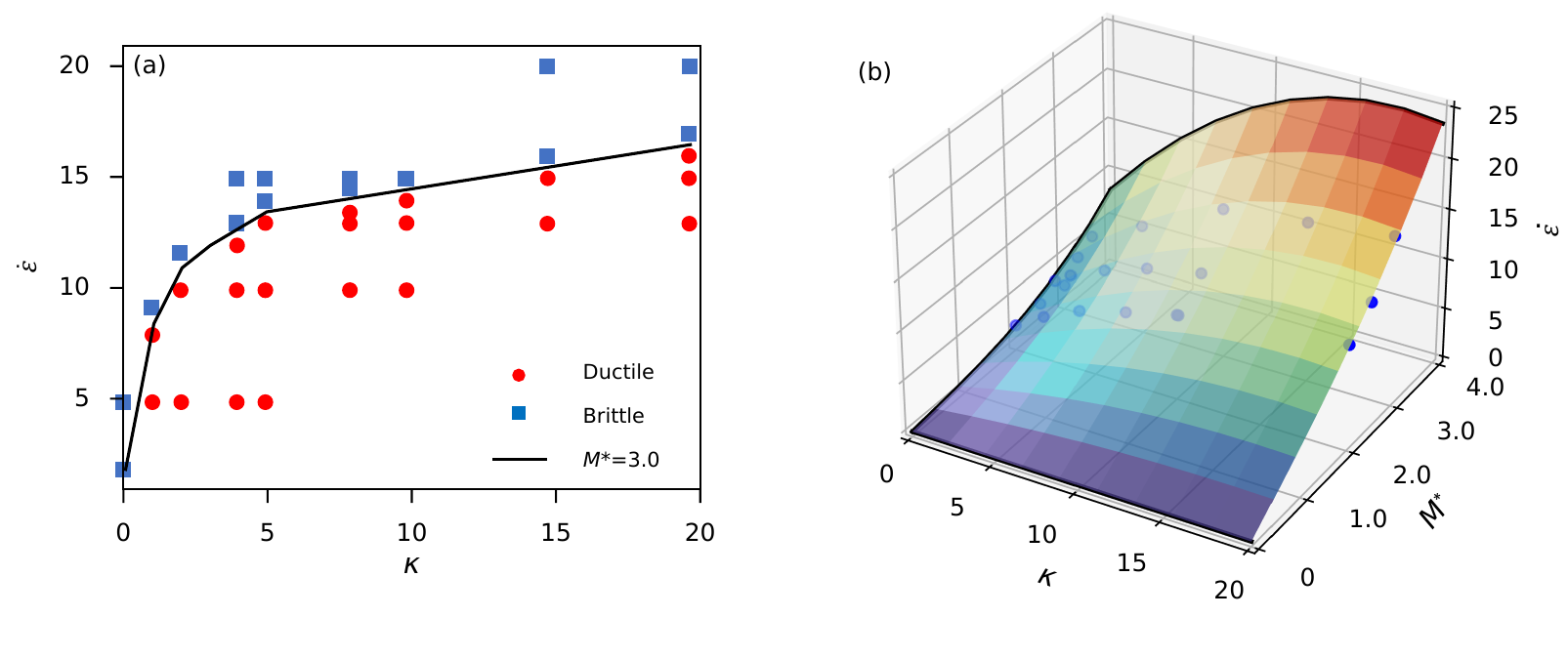}\hspace{-1em}%
\caption{\label{fracture}
(a) the black line denotes the transition strain rate line between ductile and brittle fracture when the interfacial strength is 3.0, where the red dot and blue squares represent the ductile and brittle fracture respectively; 
(b) the transition boundary between ductile and brittle fracture is denoted by the colored surface, where the blue dots represent the transition strain rate under different interface strength $M^*$ and reaction constant $\kappa$.
}
\end{figure}
Now we investigate the transition of ductile and brittle fracture by varying applied strain rates $\dot{\epsilon}$ and reaction constants $\kappa$. 
Each phase has single slip system, with $\theta^{\alpha} = 30^\circ$ and $\theta^\beta = -60^\circ$, respectively. 
The applied tensile force was along the $x$-axis, and the corresponding loading rate was $\dot{\sigma} = E (\dot{\epsilon} -\dot{\epsilon}^{\txp}_{xx})$. 
Other parameters are the same as in the previous cases (see Table~\ref{table:1}).
Under a fixed reaction constant $\kappa$, the materials exhibited a transition from ductile to brittle fracture as the applied strain rate increased. 
This indicates the existence of a transition strain rate corresponding to the shift in fracture behavior under a given reaction constant.

When the interface decohesion strength is 3.0 ($M^*=3.0$), Fig.~\ref{fracture}(a) illustrates the ductile fracture mode (represented by red dots) and brittle fracture mode (represented by blue squares) under different strain rates $\dot{\epsilon}$ and reaction constants $\kappa$. 
The black solid line represents the boundary between ductile and brittle fracture based on the applied strain rate. 
It can be observed that a higher applied transition strain rate is required as the reaction constant $\kappa$ increases. 
Moreover, the required transition strain rate varies when considering materials with different interface strengths $M^*$. 
The transition strain rate line corresponding to different interface strengths $M^*$ can be obtained following the same method. 
Based on the three different transition lines, the transition surface can be interpolated as shown in Fig.~\ref{fracture}(b). 
As depicted in Fig.~\ref{fracture}(b), the corresponding transition strain rate increases with an increase in interface strength $M^*$.

\subsection{Variation of strain hardening rate with strain}\label{discussion_theta}
In our simulation, the strain hardening arises solely from the back stress induced by the accumulated dislocations at the interface.
As a result, the strain hardening rate versus strain curve should exhibit significant variations under different reaction constants. 
Now we continue to investigate the strain hardening rate up-turn by this two-dimensional CDD model and the proposed interface BC.
In these cases, each phase has one slip system, with $\theta^{\alpha} = 30^\circ$ and $\theta^\beta = -60^\circ$, respectively.
The applied tensile force is along the $x$-axis, and the other parameters are the same as in the previous cases (see Table~\ref{table:1}).
\begin{figure*}[!ht]
\centering
\includegraphics[width=1.0\linewidth]{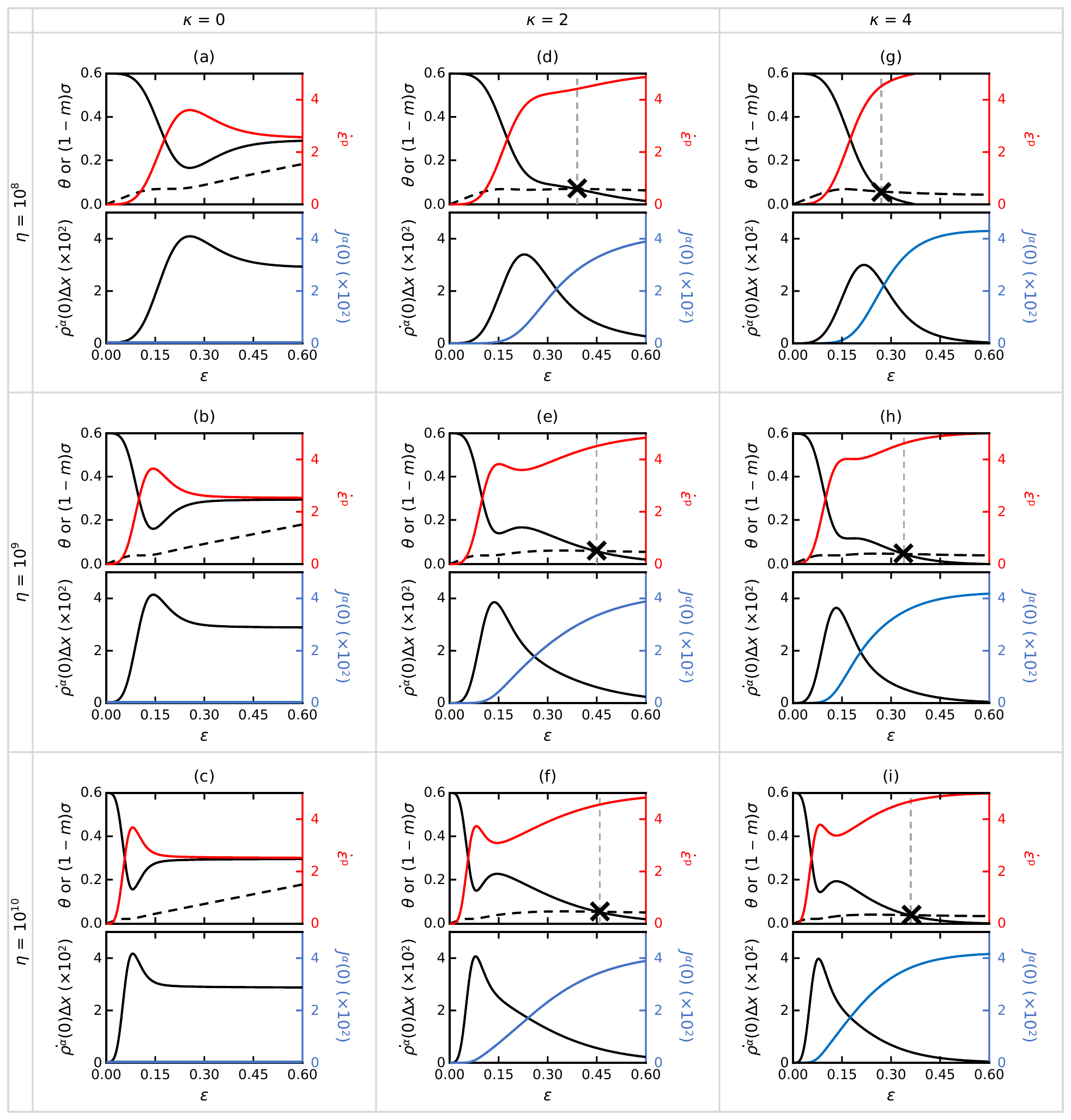}\hspace{-1em}%
\caption{\label{shape_strain_harening}
When the dislocation source intensity $\eta=10^8$, the first row (a), (d), and (g) show the evolution of strain hardening rate $\Theta$, plastic strain rate $\dot{\epsilon}^{\text{p}}$, the rate of piled up dislocation $\dot{\rho}(0) \Delta x$, and dislocation flux for $\alpha$ phase $J^{\alpha}(0)$ for three different reaction constants $\kappa=0, \ 2.0, \text{and} \ 4.0$ respectively.
When the dislocation source intensity are $\eta=10^9$ or $10^{10}$, the evolution of corresponding quantities are presented as the second and third rows respectively.
The symbol ``$\times$'' represents the necking point of the materials in each subplot.
}
\end{figure*}

\begin{figure*}[!ht]
\centering
\includegraphics[width=1.0\linewidth]{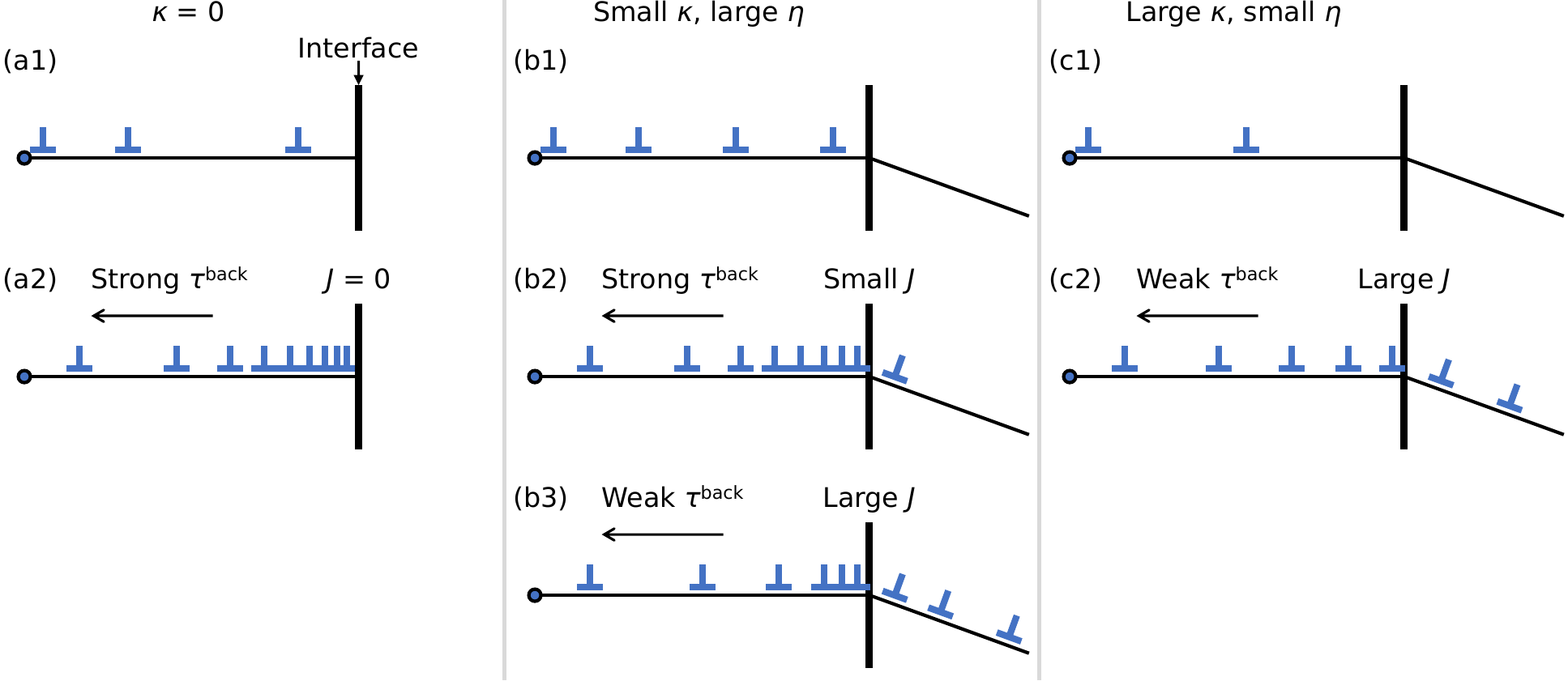}\hspace{-1em}%
\caption{\label{strain_hardening_reaction}
(a1)-(a2), (b1)-(b3), and (c1)-(c2) show the process of dislocation accumulation and transmission near the interface with the increase of reaction constant $\kappa$ respectively.
}
\end{figure*}

For a case with reaction constant of $\kappa=2.0$ and dislocation source intensity $\eta=10^9$, the strain hardening process can be classified into three stages as shown in Fig.~\ref{shape_strain_harening}(e). 
Initially, dislocations glide along the slip plane, causing plastic deformation of the material and leading to a decrease in $\Theta$. 
As dislocations pile up against the interface, the resulting back stress contributes to a transient increase in strain hardening. 
During this stage, the dislocation flux remains small, but dislocations accumulate rapidly. 
When a sufficient quantity of dislocations has piled up, the dislocation fluxes reacting at the interface increase significantly due to the penetrable/Robin interface boundary condition (Eq.~\eqref{rbc10}), leading to a decrease in the pile up and strain hardening rate. 
This whole process is illustrated schematically in Fig.~\ref{strain_hardening_reaction}(b1)-(b3).
To elucidate it more clearly, the rate of piled up dislocation at the interface is defined as $\dot{\rho}(0) \Delta x \Delta y / \Delta y = \dot{\rho}(0) \Delta x$, where $\Delta x$ and $\Delta y$ are the lengths of each mesh size. 
The corresponding dislocation flux $J^{\alpha}(0)$ and piled up rate $\dot{\rho}^{\alpha}(0) \Delta x$ are also plotted for each case.

In the case of a larger reaction constant, $\kappa=4.0$, dislocations can react quickly once they reach the interface as shown in Fig.~\ref{strain_hardening_reaction}(c1)-(c2). 
Consequently, only a small number of dislocations pile up, resulting in limited back stress and a slower increase in strain hardening. 
The strain hardening rate during the loading process does not exhibit a pronounced $\Theta$-upturn, as shown in Fig.~\ref{shape_strain_harening}(h). 
In the extreme case where the interface BC is impenetrable ($\kappa=0$), all dislocations are blocked from piling up against the interface as schematically shown in Fig.~\ref{strain_hardening_reaction}(a1)-(a2). 
This leads to an initial decrease and subsequent constant strain hardening rate throughout the loading process, as depicted in Fig.~\ref{shape_strain_harening}(b). 
In this scenario, the theoretical ductility is quite strong, as evidenced by the constant strain hardening rate denoted by the solid black line in Fig.~\ref{shape_strain_harening}(b). 
It should be noted that in experiments, typically only the decrease in strain hardening can be observed, as materials often undergo brittle fracture before reaching the stage of constant strain hardening. 
Hence, there is a competition between the rate of dislocation pile-up progress and the dislocation reaction rate at the interface, as shown in Fig.~\ref{strain_hardening_reaction}(b1)-(b3) and (c1)-(c2). 
When the source intensity $\eta$ increases while keeping the reaction constant the same, the phenomenon of $\Theta$-upturn gradually becomes more pronounced due to the dominance of the dislocation accumulation rate in these cases as shown in Fig.~\ref{shape_strain_harening}(d)-(f) or (g)-(i).
On the contrary, the $\Theta$-upturn gradually vanishes, and even the stress softening of the materials occurs when the source intensity $\eta=10^8$ as Fig.~\ref{shape_strain_harening}(d) and (g) show.

\begin{figure*}[!ht]
\centering
\includegraphics[width=0.6\linewidth]{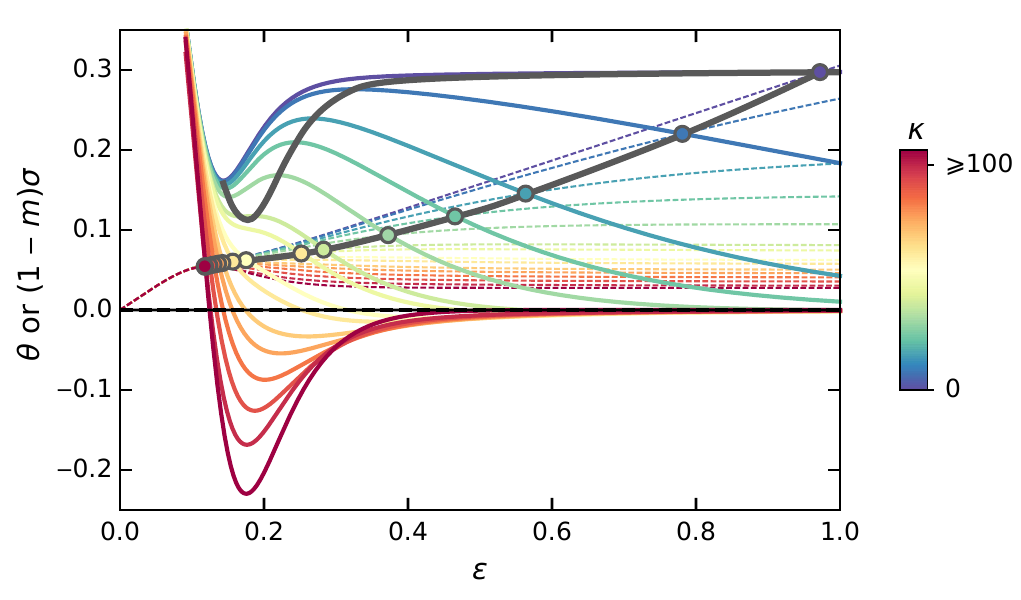}\hspace{-1em}%
\caption{\label{kappa_strain_hardening}
The strain hardening rate $\Theta$ (solid curves) and $(1-m)\sigma$ (dotted curves) are plotted under different reaction constants $\kappa$ respectively.
Instability points are indicated by circular symbols.
}
\end{figure*}

As depicted in Fig.~\ref{kappa_strain_hardening}, the $\Theta$-upturn becomes less and less obvious and disappears finally as the reaction constant $\kappa$ increases.
It should be mentioned that the materials undergo softening, i.e., the strain hardening rate gets negative, when the reaction constant is large enough.
The ductile fracture of the material is significantly delayed by the strain hardening up-turn resulting from the strong back stress induced by the piled up dislocations as the solid lines show in Fig~\ref{kappa_strain_hardening}. 
Therefore, our results qualitatively demonstrate that an appropriate reaction constant is necessary to maintain good ductility in materials. 
In other words, on one hand, the reaction constant should not be too large, as this would lead to weak strain hardening and ductile fracture. 
On the other hand, the reaction constant cannot be too small, such as in the case of impenetrable GBs, as this would result in a rapid increase in local stress and subsequent brittle fracture near the interface.

\section{Conclusions}
In this paper, we applied the mesoscale interface boundary condition to a two-dimensional continuum dislocation dynamics model. 
To validate our model, we implemented several special examples. 
Under the impenetrable boundary condition, dislocations pile up against the grain boundaries. 
Our results show that dislocations accumulate in a small region near the grain boundaries and decay quickly away from them. 
By meshing the simulation cell differently, we observed that the total amount of dislocations near the grain boundaries remains constant while the magnitude of dislocation density increases with a decrease in mesh size. 
For the point source case, the distribution of dislocation density matches the theoretical piled up dislocation distribution. 
Based on this model, we investigated the mechanical properties of materials as follows:
\begin{itemize}
\item[(i)] 
For fixed applied strain rate $\dot{\epsilon}$ and interfacial strength $M^*$, as the reaction constant $\kappa$ increases, the failure of materials is initially dominated by brittle fracture and then shifted to ductile fracture for a single slip system within each phase. 
Therefore, there is a transition reaction constant $\kappa_{\text{T}}$ from brittle to ductile fracture mode.
Besides, with the varying reaction constant, there is a critical $\kappa^*$ corresponding to the peak value of ductility.
When the reaction constant is below the critical point $\kappa^*$, the ductility of the materials increases because more dislocations can react at the interface and cause plastic deformation. 
However, when the reaction constant continues to increase and surpasses the critical point $\kappa^*$, the ductility gradually decreases due to weak strain hardening. 
This is because, in this two-dimensional model, we only consider the strengthening mechanism arising from grain boundary strengthening, while other mechanisms such as Taylor hardening between different slip systems or precipitation hardening within the grains are not taken into consideration.
\item[(ii)] 
Under a fixed reaction constant $\kappa$, the fracture mode can shift from ductile to brittle fracture mode with an increase in the applied strain rate $\dot{\epsilon}$. 
Therefore, there exists a transition applied strain rate that distinguishes the ductile and brittle fracture domains under different reaction constants. 
Based on our simulation results, a larger transition strain rate is required as the reaction constant increases. 
For materials with a weaker interface strength $M^*$, a smaller transition strain rate is required.

\item[(iii)] 
The variation of the strain hardening rate with strain can be well explained by our simple model. 
The strain hardening rate up-turn can be attributed to the competition between the rate of dislocation pileup and the dislocation reaction rate at the interface. 
The strain hardening rate initially decreases as dislocation sources are activated and dislocations glide along the slip plane within the grains. 
As dislocations pile up and the back stress strengthens, the strain hardening rate transiently increases. 
Due to the interface boundary condition, as dislocation fluxes increase and the effect of piled up dislocations weakens, the strain hardening rate decreases again. 
When the reaction constant $\kappa$ becomes large and the entire process is dominated by the dislocation reaction rate, the strain hardening and ductility become weaker. 
Thus, the strain hardening up-turn gradually diminishes as the reaction constant continues to increase, while it can be enhanced when the generation rate of dislocations increases.
\end{itemize}

\section*{Acknowledgement}
JY, DJS and JH were supported by the National Key R\&D Program of China (2021YFA1200202). 
JY, AHWN and DJS also gratefully acknowledge support of the Hong Kong Research Grants Council Collaborative Research Fund C1005-19G. 
JH acknowledges support from the Early Career Scheme (ECS) Grant of the Hong Kong Research Grants Council 21213921. 
AHWN also acknowledges support from the Shenzhen Fund 2021 Basic Research General Programme JCY20210324115400002 and the Guangdong Province Basic and Applied Research Key Project 2020B0301030001.

\appendix
\section{Elastic interaction among dislocations}\label{section_stress_calculation}
\subsection{The calculation of stress kernel}
\begin{figure}[htb]
\centering
\includegraphics[width=1\linewidth]
{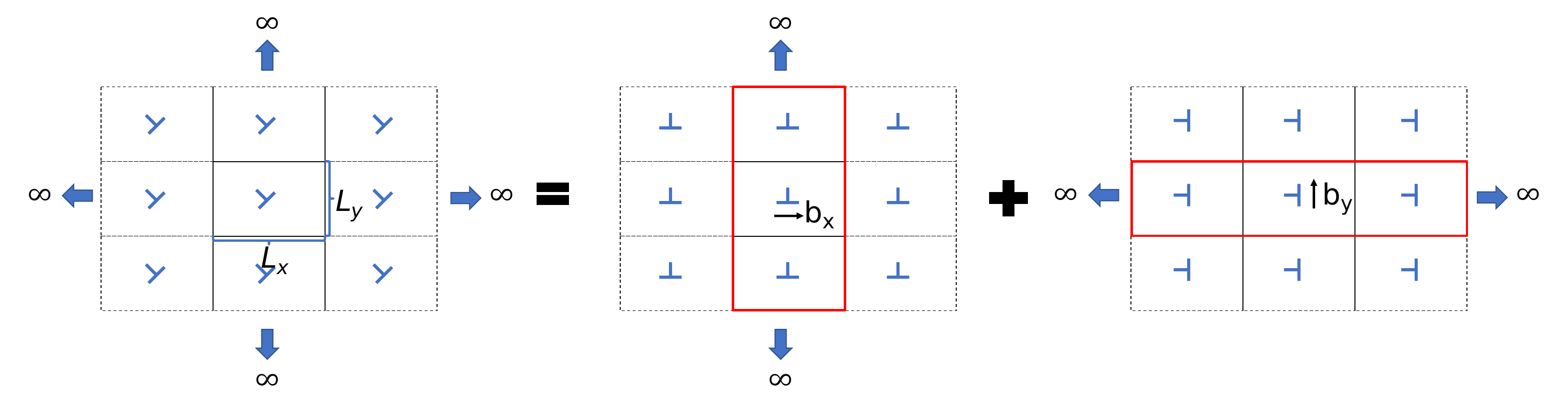}\hspace{-1.78em}%
\caption{\label{config_of_stress_kernel}
 Dislocations within two dimensional periodic simulation cells can be decomposed of two array dislocations.
}
\end{figure}

As Fig.~\ref{config_of_stress_kernel} shows, the boundary condition is a doubly periodic system in $x$ and $y$-direction.
Analytical expressions for an array of dislocations in one direction are available in classical textbook~\citep{hirth1982theory}.
Only the summation over one direction can be performed analytically, the summation over the other direction has to be performed numerically.
The array of dislocation can be decomposed into two parts: the Burgers vector $b_x$ parallel to $x$-axis and Burgers vectors $b_y$ normal to $x$-axis. 
The stress field of an infinitely long straight dislocation summed over $y$-direction with Burger vector $b_x$ can be expressed
\begin{equation}\label{tauwperp_y}
\left\{
\renewcommand*{\arraystretch}{1.2}
\begin{array}{lll}
\sigma_{xx}^{\txw\perp}
= -\sigma_0 \sin Y \left(
\cosh X - \cos Y + X\sinh X \right)
\\
\sigma_{yy}^{\txw\perp}
= -\sigma_0 \sin Y\left(
\cosh X - \cos Y - X\sinh X \right)
\\
\sigma_{xy}^{\txw\perp}
= \sigma_0 X \left(\cosh X \cos Y - 1\right)
\end{array}\right.
\end{equation}
where
\begin{equation}
\sigma_0
\equiv
\frac{K \pi b^{\perp}/L_y}{\left(
\cosh X - \cos Y
\right)^2},
\quad
X \equiv \frac{2\pi x}{L_y},
\quad
Y \equiv \frac{2\pi y}{L_y}, 
\end{equation}
where the superscript ``$\text{w}\perp$'' represents the dislocation wall with Burgers vector $b_x$ and $L_y$ is the length of unit cell in $y$-direction. 
The stress fields only have short-range stress and decay exponentially along $x$-direction. 
Therefore, the numerical summations along $x$-direction tend to converge quiet quickly.

For an array dislocations with Burgers vector $b_y$, the stress fields summed over $y$-direction analytically have long-range stress. 
The numerical summation over $x$-direction is conditional convergence because of the existence of long-range stress.
The final results of stress fields are dependent of the order of summation, i.e., the shape of truncation domain~\citep{cai2003periodic,kuykendall2013conditional}. 
For consistency, the stress fields are required to be corrected by deducting the spurious stress field.

In this paper, we solve this problem by summing the dislocations with Burgers vector $b_y$ over $x$-direction analytically.
The stress fields can be considered as rotating the stress fields of first types by $90^\circ$. The final form can be expressed as
\begin{equation}\label{tauwperp_x}
\left\{
\renewcommand*{\arraystretch}{1.2}
\begin{array}{lll}
\sigma_{xx}^{\txw\perp}
= \sigma_0 \sin X\left(
\cosh Y - \cos X - Y\sinh Y \right)
\\
\sigma_{yy}^{\txw\dashv}
= -\sigma_0 \sin X \left(
\cosh Y - \cos X + Y\sinh Y \right)
\\
\sigma_{xy}^{\txw\perp}
= -\sigma_0 Y \left(\cosh Y \cos X - 1\right)
\end{array}\right.
\end{equation}
where
\begin{equation}
\sigma_0
\equiv
\frac{K \pi b^{\dashv}/L_x}{\left(
\cosh X - \cos Y
\right)^2},
\quad
X \equiv \frac{2\pi x}{L_x},
\quad
Y \equiv \frac{2\pi y}{L_x}, 
\end{equation}
where the superscript ``$\text{w}\dashv$'' and $L_x$ ($L_x=L_\alpha+L_\beta$) represent the dislocation wall with Burgers vector $b_y$ and length of unit cell in $x$-direction respectively. 
Since these stress fields are all short-range stress, the summation of series will be absolute convergent.  

\subsection{Calculation of stress}
Once the stress kernels produced by single dislocation under doubly periodic boundary condition are obtained, we can superimpose the stress caused by all the dislocations within the whole area. 
Given all the dislocation density $\rho^{(i)}(x,y)$ of the whole area, the net dislocation density can be decomposed into two parts: 
\begin{equation}
    \rho^{\perp}(x,y) = \sum_i \rho^{(i)}(x,y) \cos\theta^{(i)}, \quad
    \rho^{\dashv}(x,y) = \sum_i \rho^{(i)}(x,y) \sin\theta^{(i)},
\end{equation}
where $\rho^{\perp}$ and $\rho^{\dashv}$ are the total net dislocation density with Burgers vector $b_x$ and $b_y$ respectively as introduced above.
Then we can update the total stress by summing the external stress and internal stress caused by all dislocations. 
The component of total stress can be expressed
\begin{equation}
    \sigma_{ij}(x,y) = \sigma_{ij}^{\text{ext}}(x,y) + \iint_{\text{cell}} \rho^{\dashv}(x',y') \sigma_{ij}^{\txw\dashv}((x'-x),(y'-y))dx'dy' + \iint_{\text{cell}} \rho^{\perp}(x',y') \sigma_{ij}^{\txw\perp}((x'-x),(y'-y))dx'dy'.
\end{equation}
Where $\sigma_{ij}^{\txw\perp}$ and $\sigma_{ij}^{\txw\dashv}$ represent the stress kernel produced by $b_x$ and $b_y$ components of single dislocation under doubly periodic boundary condition respectively.

\section{Numerical simulation}\label{numerical_simulations}
Consider a unit cell which is bounded at $x=0$ and $L_x$ along the $\mathbf{e}_x$-axis and bounded at $y=0$ and $L_y$ along the $\mathbf{e}_y$-axis as shown in Fig.~\ref{twodimension_model}. 
Discretize the $x$ and $y$ coordinate by $x_i = i\Delta x$ ($i=0,\cdots,N_x$) and $y_j = j\Delta y$ ($j=0,\cdots,N_y$) respectively, where the interval is $\Delta x = L_x/N_x$ ans $\Delta y = L_y/N_y$. 
Discretize the time $t$ by $t_n = n\Delta t$ ($n = 0, 1, \cdots$), where $\Delta t$ is the time step. 

\begin{itemize}
\item[{\bf Step 0}] Calculate the initial resolved shear stress and set the initial dislocation density
\begin{align}\label{rss}
    \tau^{(i)}(x_i,y_j,0) = \sigma_{xx}(x_i,y_j,0) s_x^{(i)} n_x^{(i)} &+\sigma_{yy}(x_i,y_j,0) s_y^{(i)} n_y^{(i)} + \sigma_{xy}(x_i,y_j,0) (s_x^{i} n_y^{(i)} + s_y^{(i)} n_x^{(i)}), \\
    \rho_{+}^{(i)}(x_i,y_j,0) &= \rho_{-}^{(i)}(x_i,y_j,0) = \rho^{(i)}_0,
\end{align}
where $\mathbf{s}^{(i)}$ and $\mathbf{n}^{(i)}$ are the slip direction and unit normal direction of $i$ slip plane respectively.
\item[{\bf Step 1}]
Calculate the velocity at each site by 
\begin{equation}
v^{(i)}_{+}(x_i,y_j,t_n)
= v_0\left(\dfrac{\tau^{(i)}(x_i,y_j,t_n)-\tau_{\text{f}}}{\tau_0}\right)^n, v^{(i)}_{-}(x_i,y_j,t_n) = - v^{(i)}_{+}(x_i,y_j,t_n)
\quad (i,j=0,\cdots,N).
\end{equation}

\item[{\bf Step 2}]
Apply one of the BCs:

Dirichlet BC: let $\rho^{(i)}_{+/-}(x_0,t_n) = \rho^{(i)}_{+/-}(x_N,t_n) = 0$;

Neumann BC: let $\rho^{(i)}_{+/-}(x_0,t_n) v^{(i)}_{+/-}(x_0,t_n) = \rho^{(i)}_{+/-}(x_N,t_n) v^{(i)}_{+/-}(x_N,t_n) = 0$;

Robin BC: let $\rho^{(i)}_{+/-}(x_0,t_n)v^{(i)}_{+/-}(x_0,t_n) = \rho^{(i)}_{+/-}(x_N,t_n)v^{(i)}_{+/-}(x_N,t_n) =J^{(i)}_{+/-}$. 

\item[{\bf Step 3}]
Integrate the differential equations by the finite difference method. 
First, the net source which is the sum of the generation term and the annihilation terms can be expressed as 
\begin{equation}
    \dot{\rho}^{(i),\text{source}}(x_i, y_j,t_n) = -r_{\text{c}} \rho_{+}^{(i)}(x_i,y_j,t_n) \rho_{-}^{(i)} (x_i,y_j,t_n)\left|v_{+}^{(i)}(x_i,y_j,t_n) - v_{-}^{(i)}(x_i,y_j,t_n)\right| +\eta (\tau^{(i)}(x_i,y_j,t_n))^m.
\end{equation}
The new positive/negative dislocation densities at each site are updated by the evolution equations~\eqref{evolution_of_dd}.
In order to handle high gradients, a total variation diminishing scheme must be introduced, usually Monotone Upstream-centred Schemes for Conservation laws (MUSCL) are employed. Here, we used the Kurganov–Tadmor central scheme, which is a second-order, high-resolution MUSCL construction.
Since it is applicable to all slip systems within each grain, for the sake of brevity, the slip system denoted by subscript ``$(i)$'' and subscript ``$+/-$'' sign for positive/negative dislocations are omitted here.
With this, the flux operator above is re-cast into the semi-discrete form~\citep{wesseling2009principles}:
\begin{equation}
    \frac{d\rho_i}{dt} + \frac{1}{\Delta x_i}[(\rho v)_{i+1/2} - (\rho v)_{i-1/2}] + \cdots
\end{equation}
where:
\begin{align}
 &\rho_i \equiv \rho(x_i,t_n);\quad
 (\rho v)_{i \pm 1/2} = \frac{1}{2}{[\rho^R_{i \pm 1/2} v_{i \pm 1/2} - \rho^L_{i \pm 1/2} v_{i \pm 1/2}] - \mathrm{max}[|v_i|,|v_{i \pm 1}|][\rho^R - \rho^L]};\\
 &\rho^L_{i + 1/2} = \rho_i + 0.5 \Phi(r_i)(\rho_{i+1} - \rho_i); \quad
 \rho^R_{i + 1/2} = \rho_i - 0.5 \Phi(r_{i+1})(\rho_{i+2} - \rho_{i+1});\\
&\rho^L_{i - 1/2} = \rho_{i-1} + 0.5 \Phi(r_{i-1})(\rho_{i} - \rho_{i-1}); \quad
 \rho^R_{i - 1/2} = \rho_i - 0.5 \Phi(r_{i})(\rho_{i+1} - \rho_{i});\\
 & r_i = \frac{\rho_i - \rho_{i-1}}{\rho_{i+1} - \rho_i}; \quad
 \Phi(r) = \frac{2r}{1+r^2};
\end{align}
Here, $\Phi(r)$ is the flux limiter, which is a function such that for $r<0$,$\Phi= 0$ and $\Phi= 1$ for $r= 1$. The main idea of MUCSL is to use slope limited left and right extrapolated states for the cell’s approximation, and the left and right states are those at the cell walls, i.e., $i= \pm 1/2$. Also MUSCL is combined with the Runge–Kutta time marching method to obtain accurate results.
\item[{\bf Step 4}]Calculate the net dislocation density:
\begin{equation}
    \rho^{(i)}(x_i,y_i,t_n)=\rho^{(i)}_+(x_i,y_i,t_n) - \rho^{(i)}_-(x_i,y_i,t_n),
\end{equation}
The dissociation of dislocation densities with respect to Burgers vector $b_x$ and $b_y$ are 
\begin{equation}
    \rho^{\dashv}(x_i,y_i,t_n)=\sum_{i} \rho^{(i)}(x_i,y_i,t_n) \cos\theta^{(i)}, \quad 
    \rho^{\perp}(x_i,y_i,t_n)=\sum_{i} \rho^{(i)}(x_i,y_i,t_n) \sin\theta^{(i)}.
\end{equation}
\item[{\bf Step 5}] Update the stress field of the model
\begin{align}
    \sigma_{kl}(x_i, y_j, t_{n+1}) = &\sigma_{kl}^{\text{ext}}(x_i, y_j,t_n) + \sum_{i'=0}^{N_x} \sum_{j'=0}^{N_y} \rho^{\dashv}(x_{i'}, y_{j'},t_n) \sigma_{kl}^{\txw\dashv}((i'-i) \Delta x,(j'-j)\Delta y)\Delta x \Delta y + \\
    &\sum_{i'=0}^{N_x} \sum_{j'=0}^{N_y} \rho^{\perp}(x_{i'}, y_{j'},t_n) \sigma_{kl}^{\txw\perp}((i'-i)\Delta x,(j'-j)\Delta y)\Delta x \Delta y,
\end{align}
where $\sigma_{kl}$ represents $\sigma_{xy}$, $\sigma_{xx}$ and $\sigma_{yy}$,  $\sigma_{kl}^{\txw\dashv}$ and $\sigma_{kl}^{\txw\perp}$ are the corresponding stress kernel as Eq.~\eqref{tauwperp_y} and Eq.~\eqref{tauwperp_x} show. 
\item[{\bf Step 6}] Update the resolved shear stress by Eq.~\eqref{rss}, then go to {\bf Step 1}.
\end{itemize}

\begin{table}[htb]
\centering
\caption{Non-dimensional parameters for the two-dimensional continuous dislocation dynamics simulations. 
For simplicity, we choose the same parameters for all slip systems with two phases (including the interface).}
\hspace{4cm}
\renewcommand{\arraystretch}{1.1} 
\begin{tabular}{m{10cm} m{2cm}} 
 \hline
 Parameters & Values \\ [1ex] 
 \hline
 Burgers vector magnitude,  $b$ & $1 \times 10^{-3}$ \\ 
 Annihilation capture radius,  $r_{\text{c}}$ & $1 \times 10^{-3}$ \\
Time step, $\Delta t$  & $3 \times 10^{-5}$ \\ 
Length of unit cell along $x$ axis, $L_x$ & 1.0 \\
Exponent in the velocity power law, $n$ & 1 \\
Slip resistance, $\tau_0$ & 0.1 \\
Peierls stress, $\tau_{\text{f}}$ & $1 \times 10^{-2}$ \\
Source intensity, $\eta$ & $1 \times 10^8$ \\
Source term exponent,  $m$ & 2 \\ [1ex] 
 \hline
\end{tabular}
\label{table:1}
\end{table}

\bibliographystyle{elsarticle-harv}
\bibliography{mybib}

\providecommand{\noopsort}[1]{}\providecommand{\singleletter}[1]{#1}%
\begin{thebibliography}{43}
\expandafter\ifx\csname natexlab\endcsname\relax\def\natexlab#1{#1}\fi
\providecommand{\url}[1]{\texttt{#1}}
\providecommand{\href}[2]{#2}
\providecommand{\path}[1]{#1}
\providecommand{\DOIprefix}{doi:}
\providecommand{\ArXivprefix}{arXiv:}
\providecommand{\URLprefix}{URL: }
\providecommand{\Pubmedprefix}{pmid:}
\providecommand{\doi}[1]{\href{http://dx.doi.org/#1}{\path{#1}}}
\providecommand{\Pubmed}[1]{\href{pmid:#1}{\path{#1}}}
\providecommand{\bibinfo}[2]{#2}
\ifx\xfnm\relax \def\xfnm[#1]{\unskip,\space#1}\fi
\bibitem[{Cai et~al.(2003)Cai, Bulatov, Chang, Li and Yip}]{cai2003periodic}
\bibinfo{author}{Cai, W.}, \bibinfo{author}{Bulatov, V.V.},
  \bibinfo{author}{Chang, J.}, \bibinfo{author}{Li, J.}, \bibinfo{author}{Yip,
  S.}, \bibinfo{year}{2003}.
\newblock \bibinfo{title}{Periodic image effects in dislocation modelling}.
\newblock \bibinfo{journal}{Philosophical Magazine} \bibinfo{volume}{83},
  \bibinfo{pages}{539--567}.
\newblock \DOIprefix\doi{https://doi.org/10.1080/0141861021000051109}.
\bibitem[{Cai and Nix(2018)}]{cai2018imperfections}
\bibinfo{author}{Cai, W.}, \bibinfo{author}{Nix, W.D.}, \bibinfo{year}{2018}.
\newblock \bibinfo{title}{Imperfections in crystalline solids}.
\bibitem[{Cleveringa and Needleman(1997)}]{cleveringa1997comparison}
\bibinfo{author}{Cleveringa, H.}, \bibinfo{author}{Needleman, A.},
  \bibinfo{year}{1997}.
\newblock \bibinfo{title}{Comparison of discrete dislocation and continuum
  plasticity predictions for a composite material}.
\newblock \bibinfo{journal}{Acta Materialia} \bibinfo{volume}{45},
  \bibinfo{pages}{3163--3179}.
\newblock \DOIprefix\doi{https://doi.org/10.1016/S1359-6454(97)00011-6}.
\bibitem[{Cottrell(1965)}]{cottrell1965mechanics}
\bibinfo{author}{Cottrell, A.H.}, \bibinfo{year}{1965}.
\newblock \bibinfo{title}{Mechanics of fracture in large structures}.
\newblock \bibinfo{journal}{Proceedings of the Royal Society of London. Series
  A. Mathematical and Physical Sciences} \bibinfo{volume}{285},
  \bibinfo{pages}{10--21}.
\newblock \DOIprefix\doi{https://doi.org/10.1098/rspa.1965.0085}.
\bibitem[{Dickel et~al.(2014)Dickel, Schulz, Schmitt and
  Gumbsch}]{dickel2014dipole}
\bibinfo{author}{Dickel, D.}, \bibinfo{author}{Schulz, K.},
  \bibinfo{author}{Schmitt, S.}, \bibinfo{author}{Gumbsch, P.},
  \bibinfo{year}{2014}.
\newblock \bibinfo{title}{Dipole formation and yielding in a two-dimensional
  continuum dislocation model}.
\newblock \bibinfo{journal}{Physical Review B} \bibinfo{volume}{90},
  \bibinfo{pages}{094118}.
\newblock \DOIprefix\doi{https://doi.org/10.1103/PhysRevB.90.094118}.
\bibitem[{El-Awady et~al.(2009)El-Awady, Wen and Ghoniem}]{el2009role}
\bibinfo{author}{El-Awady, J.A.}, \bibinfo{author}{Wen, M.},
  \bibinfo{author}{Ghoniem, N.M.}, \bibinfo{year}{2009}.
\newblock \bibinfo{title}{The role of the weakest-link mechanism in controlling
  the plasticity of micropillars}.
\newblock \bibinfo{journal}{Journal of the Mechanics and Physics of Solids}
  \bibinfo{volume}{57}, \bibinfo{pages}{32--50}.
\newblock \DOIprefix\doi{https://doi.org/10.1016/j.jmps.2008.10.004}.
\bibitem[{Hall(1951)}]{hall1951deformation}
\bibinfo{author}{Hall, E.}, \bibinfo{year}{1951}.
\newblock \bibinfo{title}{The deformation and ageing of mild steel: Iii
  discussion of results}.
\newblock \bibinfo{journal}{Proceedings of the Physical Society. Section B}
  \bibinfo{volume}{64}, \bibinfo{pages}{747}.
\newblock \DOIprefix\doi{10.1088/0370-1301/64/9/303}.
\bibitem[{Hart(1967)}]{hart1967theory}
\bibinfo{author}{Hart, E.}, \bibinfo{year}{1967}.
\newblock \bibinfo{title}{Theory of the tensile test}.
\newblock \bibinfo{journal}{Acta metallurgica} \bibinfo{volume}{15},
  \bibinfo{pages}{351--355}.
\newblock \DOIprefix\doi{https://doi.org/10.1016/0001-6160(67)90211-8}.
\bibitem[{Hirth(1972)}]{hirth1972influence}
\bibinfo{author}{Hirth, J.P.}, \bibinfo{year}{1972}.
\newblock \bibinfo{title}{The influence of grain boundaries on mechanical
  properties}.
\newblock \bibinfo{journal}{Metallurgical Transactions} \bibinfo{volume}{3},
  \bibinfo{pages}{3047--3067}.
\newblock \DOIprefix\doi{https://doi.org/10.1007/BF02661312}.
\bibitem[{Hirth and Lothe(1982)}]{hirth1982theory}
\bibinfo{author}{Hirth, J.P.}, \bibinfo{author}{Lothe, J.},
  \bibinfo{year}{1982}.
\newblock \bibinfo{title}{Theory of dislocations}.
\newblock \bibinfo{publisher}{John Wiley \& Sons}.
\bibitem[{Huang et~al.(2018)Huang, Wang, Ma, Yin, H{\"o}ppel, G{\"o}ken, Wu,
  Gao and Zhu}]{huang2018interface}
\bibinfo{author}{Huang, C.}, \bibinfo{author}{Wang, Y.}, \bibinfo{author}{Ma,
  X.}, \bibinfo{author}{Yin, S.}, \bibinfo{author}{H{\"o}ppel, H.},
  \bibinfo{author}{G{\"o}ken, M.}, \bibinfo{author}{Wu, X.},
  \bibinfo{author}{Gao, H.}, \bibinfo{author}{Zhu, Y.}, \bibinfo{year}{2018}.
\newblock \bibinfo{title}{Interface affected zone for optimal strength and
  ductility in heterogeneous laminate}.
\newblock \bibinfo{journal}{Materials Today} \bibinfo{volume}{21},
  \bibinfo{pages}{713--719}.
\newblock \DOIprefix\doi{https://doi.org/10.1016/j.mattod.2018.03.006}.
\bibitem[{Javaid et~al.(2021)Javaid, Pouriayevali and
  Durst}]{javaid2021dislocation}
\bibinfo{author}{Javaid, F.}, \bibinfo{author}{Pouriayevali, H.},
  \bibinfo{author}{Durst, K.}, \bibinfo{year}{2021}.
\newblock \bibinfo{title}{Dislocation--grain boundary interactions: Recent
  advances on the underlying mechanisms studied via nanoindentation testing}.
\newblock \bibinfo{journal}{Journal of Materials Research}
  \bibinfo{volume}{36}, \bibinfo{pages}{2545--2557}.
\newblock \DOIprefix\doi{https://doi.org/10.1557/s43578-020-00096-z}.
\bibitem[{Ji et~al.(2023)Ji, Zhou, Vivegananthan, Wu, Gao and
  Zhou}]{ji2023recent}
\bibinfo{author}{Ji, W.}, \bibinfo{author}{Zhou, R.},
  \bibinfo{author}{Vivegananthan, P.}, \bibinfo{author}{Wu, M.S.},
  \bibinfo{author}{Gao, H.}, \bibinfo{author}{Zhou, K.}, \bibinfo{year}{2023}.
\newblock \bibinfo{title}{Recent progress in gradient-structured metals and
  alloys}.
\newblock \bibinfo{journal}{Progress in Materials Science} ,
  \bibinfo{pages}{101194}\DOIprefix\doi{https://doi.org/10.1016/j.pmatsci.2023.101194}.
\bibitem[{Jiang et~al.(2019)Jiang, Devincre and Monnet}]{jiang2019effects}
\bibinfo{author}{Jiang, M.}, \bibinfo{author}{Devincre, B.},
  \bibinfo{author}{Monnet, G.}, \bibinfo{year}{2019}.
\newblock \bibinfo{title}{Effects of the grain size and shape on the flow
  stress: A dislocation dynamics study}.
\newblock \bibinfo{journal}{International Journal of Plasticity}
  \bibinfo{volume}{113}, \bibinfo{pages}{111--124}.
\newblock \DOIprefix\doi{https://doi.org/10.1016/j.ijplas.2018.09.008}.
\bibitem[{Johnston and Gilman(1959)}]{johnston1959dislocation}
\bibinfo{author}{Johnston, W.G.}, \bibinfo{author}{Gilman, J.J.},
  \bibinfo{year}{1959}.
\newblock \bibinfo{title}{Dislocation velocities, dislocation densities, and
  plastic flow in lithium fluoride crystals}.
\newblock \bibinfo{journal}{Journal of Applied Physics} \bibinfo{volume}{30},
  \bibinfo{pages}{129--144}.
\newblock \DOIprefix\doi{https://doi.org/10.1063/1.1735121}.
\bibitem[{Kacher et~al.(2014)Kacher, Eftink, Cui and
  Robertson}]{kacher2014dislocation}
\bibinfo{author}{Kacher, J.}, \bibinfo{author}{Eftink, B.},
  \bibinfo{author}{Cui, B.}, \bibinfo{author}{Robertson, I.},
  \bibinfo{year}{2014}.
\newblock \bibinfo{title}{Dislocation interactions with grain boundaries}.
\newblock \bibinfo{journal}{Current Opinion in Solid State and Materials
  Science} \bibinfo{volume}{18}, \bibinfo{pages}{227--243}.
\newblock \DOIprefix\doi{https://doi.org/10.1016/j.cossms.2014.05.004}.
\bibitem[{Kacher and Robertson(2012)}]{kacher2012quasi}
\bibinfo{author}{Kacher, J.}, \bibinfo{author}{Robertson, I.},
  \bibinfo{year}{2012}.
\newblock \bibinfo{title}{Quasi-four-dimensional analysis of dislocation
  interactions with grain boundaries in 304 stainless steel}.
\newblock \bibinfo{journal}{Acta Materialia} \bibinfo{volume}{60},
  \bibinfo{pages}{6657--6672}.
\newblock \DOIprefix\doi{https://doi.org/10.1016/j.actamat.2012.08.036}.
\bibitem[{Kacher and Robertson(2014)}]{kacher2014situ}
\bibinfo{author}{Kacher, J.}, \bibinfo{author}{Robertson, I.M.},
  \bibinfo{year}{2014}.
\newblock \bibinfo{title}{In situ and tomographic analysis of dislocation/grain
  boundary interactions in $\alpha$-titanium}.
\newblock \bibinfo{journal}{Philosophical Magazine} \bibinfo{volume}{94},
  \bibinfo{pages}{814--829}.
\newblock \DOIprefix\doi{https://doi.org/10.1080/14786435.2013.868942}.
\bibitem[{Kheradmand et~al.(2010)Kheradmand, Barnoush and
  Vehoff}]{kheradmand2010investigation}
\bibinfo{author}{Kheradmand, N.}, \bibinfo{author}{Barnoush, A.},
  \bibinfo{author}{Vehoff, H.}, \bibinfo{year}{2010}.
\newblock \bibinfo{title}{Investigation of the role of grain boundary on the
  mechanical properties of metals}, in: \bibinfo{booktitle}{Journal of Physics:
  Conference Series}, \bibinfo{organization}{IOP Publishing}. p.
  \bibinfo{pages}{012017}.
\newblock \DOIprefix\doi{10.1088/1742-6596/240/1/012017}.
\bibitem[{Kocks et~al.(1975)Kocks, Argon and Ashby}]{kocks1975progress}
\bibinfo{author}{Kocks, U.F.}, \bibinfo{author}{Argon, A.S.},
  \bibinfo{author}{Ashby, M.F.}, \bibinfo{year}{1975}.
\newblock \bibinfo{title}{Kinetics}, in: \bibinfo{booktitle}{Thermodynamics and
  kinetics of slip}. volume~\bibinfo{volume}{19}. chapter~\bibinfo{chapter}{3},
  pp. \bibinfo{pages}{68--109}.
\bibitem[{Kuykendall and Cai(2013)}]{kuykendall2013conditional}
\bibinfo{author}{Kuykendall, W.P.}, \bibinfo{author}{Cai, W.},
  \bibinfo{year}{2013}.
\newblock \bibinfo{title}{Conditional convergence in two-dimensional
  dislocation dynamics}.
\newblock \bibinfo{journal}{Modelling and Simulation in Materials Science and
  Engineering} \bibinfo{volume}{21}, \bibinfo{pages}{055003}.
\newblock \DOIprefix\doi{10.1088/0965-0393/21/5/055003}.
\bibitem[{Leung et~al.(2015)Leung, Leung, Cheng and Ngan}]{leung2015new}
\bibinfo{author}{Leung, H.S.}, \bibinfo{author}{Leung, P.S.S.},
  \bibinfo{author}{Cheng, B.}, \bibinfo{author}{Ngan, A.H.W.},
  \bibinfo{year}{2015}.
\newblock \bibinfo{title}{A new dislocation-density-function dynamics scheme
  for computational crystal plasticity by explicit consideration of dislocation
  elastic interactions}.
\newblock \bibinfo{journal}{International Journal of Plasticity}
  \bibinfo{volume}{67}, \bibinfo{pages}{1--25}.
\newblock \DOIprefix\doi{https://doi.org/10.1016/j.ijplas.2014.09.009}.
\bibitem[{Li et~al.(2010)Li, Wei, Lu, Lu and Gao}]{li2010dislocation}
\bibinfo{author}{Li, X.}, \bibinfo{author}{Wei, Y.}, \bibinfo{author}{Lu, L.},
  \bibinfo{author}{Lu, K.}, \bibinfo{author}{Gao, H.}, \bibinfo{year}{2010}.
\newblock \bibinfo{title}{Dislocation nucleation governed softening and maximum
  strength in nano-twinned metals}.
\newblock \bibinfo{journal}{Nature} \bibinfo{volume}{464},
  \bibinfo{pages}{877--880}.
\newblock \DOIprefix\doi{https://doi.org/10.1038/nature08929}.
\bibitem[{Liu et~al.(2023)Liu, Cheng, Sui, Fu and Duan}]{liu2023microstructure}
\bibinfo{author}{Liu, W.}, \bibinfo{author}{Cheng, Y.}, \bibinfo{author}{Sui,
  H.}, \bibinfo{author}{Fu, J.}, \bibinfo{author}{Duan, H.},
  \bibinfo{year}{2023}.
\newblock \bibinfo{title}{Microstructure-based intergranular fatigue crack
  nucleation model: Dislocation transmission versus grain boundary cracking}.
\newblock \bibinfo{journal}{Journal of the Mechanics and Physics of Solids}
  \bibinfo{volume}{173}, \bibinfo{pages}{105233}.
\newblock \DOIprefix\doi{https://doi.org/10.1016/j.jmps.2023.105233}.
\bibitem[{Lu et~al.(2004)Lu, Shen, Chen, Qian and Lu}]{lu2004ultrahigh}
\bibinfo{author}{Lu, L.}, \bibinfo{author}{Shen, Y.}, \bibinfo{author}{Chen,
  X.}, \bibinfo{author}{Qian, L.}, \bibinfo{author}{Lu, K.},
  \bibinfo{year}{2004}.
\newblock \bibinfo{title}{Ultrahigh strength and high electrical conductivity
  in copper}.
\newblock \bibinfo{journal}{Science} \bibinfo{volume}{304},
  \bibinfo{pages}{422--426}.
\newblock \DOIprefix\doi{10.1126/science.1092905}.
\bibitem[{Malygin(1999)}]{malygin1999dislocation}
\bibinfo{author}{Malygin, G.A.}, \bibinfo{year}{1999}.
\newblock \bibinfo{title}{Dislocation self-organization processes and crystal
  plasticity}.
\newblock \bibinfo{journal}{Physics-uspekhi} \bibinfo{volume}{42},
  \bibinfo{pages}{887}.
\newblock \DOIprefix\doi{https://doi.org/10.1070/pu1999v042n09abeh000563}.
\bibitem[{Ngan(2017)}]{ngan2017dislocation}
\bibinfo{author}{Ngan, A.H.W.}, \bibinfo{year}{2017}.
\newblock \bibinfo{title}{Dislocation-density kinematics: a simple evolution
  equation for dislocation density involving movement and tilting of
  dislocations}.
\newblock \bibinfo{journal}{MRS Communications} \bibinfo{volume}{7},
  \bibinfo{pages}{583}.
\newblock \DOIprefix\doi{https://doi.org/10.1557/mrc.2017.66}.
\bibitem[{Sarfarazi and Ghosh(1987)}]{sarfarazi1987microfracture}
\bibinfo{author}{Sarfarazi, M.}, \bibinfo{author}{Ghosh, S.},
  \bibinfo{year}{1987}.
\newblock \bibinfo{title}{Microfracture in polycrystalline solids}.
\newblock \bibinfo{journal}{Engineering fracture mechanics}
  \bibinfo{volume}{27}, \bibinfo{pages}{257--267}.
\newblock \DOIprefix\doi{https://doi.org/10.1016/0013-7944(87)90144-5}.
\bibitem[{Scardia et~al.(2014)Scardia, Peerlings, Peletier and
  Geers}]{scardia2014mechanics}
\bibinfo{author}{Scardia, L.}, \bibinfo{author}{Peerlings, R.H.},
  \bibinfo{author}{Peletier, M.A.}, \bibinfo{author}{Geers, M.G.},
  \bibinfo{year}{2014}.
\newblock \bibinfo{title}{Mechanics of dislocation pile-ups: a unification of
  scaling regimes}.
\newblock \bibinfo{journal}{Journal of the Mechanics and Physics of Solids}
  \bibinfo{volume}{70}, \bibinfo{pages}{42--61}.
\newblock \DOIprefix\doi{https://doi.org/10.1016/j.jmps.2014.04.014}.
\bibitem[{Smith and Barnby(1967)}]{smith1967crack}
\bibinfo{author}{Smith, E.}, \bibinfo{author}{Barnby, J.},
  \bibinfo{year}{1967}.
\newblock \bibinfo{title}{Crack nucleation in crystalline solids}.
\newblock \bibinfo{journal}{Metal Science Journal} \bibinfo{volume}{1},
  \bibinfo{pages}{56--64}.
\newblock \DOIprefix\doi{https://doi.org/10.1179/msc.1967.1.1.56}.
\bibitem[{Stroh(1957)}]{stroh1957theory}
\bibinfo{author}{Stroh, A.N.}, \bibinfo{year}{1957}.
\newblock \bibinfo{title}{A theory of the fracture of metals}.
\newblock \bibinfo{journal}{Advances in Physics} \bibinfo{volume}{6},
  \bibinfo{pages}{418--465}.
\newblock \DOIprefix\doi{https://doi.org/10.1080/00018735700101406}.
\bibitem[{Weertman(1996)}]{weertman1996dislocation}
\bibinfo{author}{Weertman, J.}, \bibinfo{year}{1996}.
\newblock \bibinfo{title}{Dislocation based fracture mechanics}.
\newblock \bibinfo{publisher}{World Scientific}.
\bibitem[{Wesseling(2009)}]{wesseling2009principles}
\bibinfo{author}{Wesseling, P.}, \bibinfo{year}{2009}.
\newblock \bibinfo{title}{Principles of computational fluid dynamics}.
  volume~\bibinfo{volume}{29}.
\newblock \bibinfo{publisher}{Springer Science \& Business Media}.
\bibitem[{Wu et~al.(2014)Wu, Jiang, Chen, Yuan and Zhu}]{wu2014extraordinary}
\bibinfo{author}{Wu, X.}, \bibinfo{author}{Jiang, P.}, \bibinfo{author}{Chen,
  L.}, \bibinfo{author}{Yuan, F.}, \bibinfo{author}{Zhu, Y.T.},
  \bibinfo{year}{2014}.
\newblock \bibinfo{title}{Extraordinary strain hardening by gradient
  structure}.
\newblock \bibinfo{journal}{Proceedings of the National Academy of Sciences}
  \bibinfo{volume}{111}, \bibinfo{pages}{7197--7201}.
\newblock \DOIprefix\doi{https://doi.org/10.1073/pnas.1324069111}.
\bibitem[{Wu et~al.(2015)Wu, Yang, Yuan, Wu, Wei, Huang and
  Zhu}]{wu2015heterogeneous}
\bibinfo{author}{Wu, X.}, \bibinfo{author}{Yang, M.}, \bibinfo{author}{Yuan,
  F.}, \bibinfo{author}{Wu, G.}, \bibinfo{author}{Wei, Y.},
  \bibinfo{author}{Huang, X.}, \bibinfo{author}{Zhu, Y.}, \bibinfo{year}{2015}.
\newblock \bibinfo{title}{Heterogeneous lamella structure unites
  ultrafine-grain strength with coarse-grain ductility}.
\newblock \bibinfo{journal}{Proceedings of the National Academy of Sciences}
  \bibinfo{volume}{112}, \bibinfo{pages}{14501--14505}.
\newblock \DOIprefix\doi{https://doi.org/10.1073/pnas.1517193112}.
\bibitem[{Xia and El-Azab(2015)}]{xia2015computational}
\bibinfo{author}{Xia, S.}, \bibinfo{author}{El-Azab, A.}, \bibinfo{year}{2015}.
\newblock \bibinfo{title}{Computational modelling of mesoscale dislocation
  patterning and plastic deformation of single crystals}.
\newblock \bibinfo{journal}{Modelling and Simulation in Materials Science and
  Engineering} \bibinfo{volume}{23}, \bibinfo{pages}{055009}.
\newblock \DOIprefix\doi{10.1088/0965-0393/23/5/055009}.
\bibitem[{Yang et~al.(2021)Yang, Yuan, Xie, Wang, Ma and Wu}]{yang2021strain}
\bibinfo{author}{Yang, M.}, \bibinfo{author}{Yuan, F.}, \bibinfo{author}{Xie,
  Q.}, \bibinfo{author}{Wang, Y.}, \bibinfo{author}{Ma, E.},
  \bibinfo{author}{Wu, X.}, \bibinfo{year}{2021}.
\newblock \bibinfo{title}{{Strain hardening in Fe--16Mn--10Al--0.86 C--5Ni high
  specific strength steel}}, in: \bibinfo{booktitle}{Heterostructured
  Materials}. \bibinfo{publisher}{Jenny Stanford Publishing}, pp.
  \bibinfo{pages}{721--747}.
\newblock \DOIprefix\doi{https://doi.org/10.1016/j.actamat.2016.02.044}.
\bibitem[{Yu et~al.(2023)Yu, Ngan, Srolovitz and Han}]{yu2023application}
\bibinfo{author}{Yu, J.}, \bibinfo{author}{Ngan, A.H.},
  \bibinfo{author}{Srolovitz, D.J.}, \bibinfo{author}{Han, J.},
  \bibinfo{year}{2023}.
\newblock \bibinfo{title}{Application of rigorous interface boundary conditions
  in mesoscale plasticity simulations}.
\newblock \bibinfo{journal}{Modelling and Simulation in Materials Science and
  Engineering} \DOIprefix\doi{10.1088/1361-651X/ad26a0}.
\bibitem[{Yu et~al.(2025)Yu, Ngan, Srolovitz and Han}]{yu2025mesoscale}
\bibinfo{author}{Yu, J.}, \bibinfo{author}{Ngan, A.H.},
  \bibinfo{author}{Srolovitz, D.J.}, \bibinfo{author}{Han, J.},
  \bibinfo{year}{2025}.
\newblock \bibinfo{title}{Mesoscale description of interface-mediated
  plasticity}.
\newblock \bibinfo{journal}{Acta Materialia} \bibinfo{volume}{283},
  \bibinfo{pages}{120552}.
\newblock \DOIprefix\doi{https://doi.org/10.1016/j.actamat.2024.120552}.
\bibitem[{Zhang et~al.(2021)Zhang, Lu, Zhang, Tian, Kan and
  Kang}]{zhang2021dislocation}
\bibinfo{author}{Zhang, X.}, \bibinfo{author}{Lu, S.}, \bibinfo{author}{Zhang,
  B.}, \bibinfo{author}{Tian, X.}, \bibinfo{author}{Kan, Q.},
  \bibinfo{author}{Kang, G.}, \bibinfo{year}{2021}.
\newblock \bibinfo{title}{Dislocation--grain boundary interaction-based
  discrete dislocation dynamics modeling and its application to bicrystals with
  different misorientations}.
\newblock \bibinfo{journal}{Acta Materialia} \bibinfo{volume}{202},
  \bibinfo{pages}{88--98}.
\newblock \DOIprefix\doi{https://doi.org/10.1016/j.actamat.2020.10.052}.
\bibitem[{Zhu and Li(2010)}]{zhu2010ultra}
\bibinfo{author}{Zhu, T.}, \bibinfo{author}{Li, J.}, \bibinfo{year}{2010}.
\newblock \bibinfo{title}{Ultra-strength materials}.
\newblock \bibinfo{journal}{Progress in Materials Science}
  \bibinfo{volume}{55}, \bibinfo{pages}{710--757}.
\newblock \DOIprefix\doi{https://doi.org/10.1016/j.pmatsci.2010.04.001}.
\bibitem[{Zhu et~al.(2021)Zhu, Ameyama, Anderson, Beyerlein, Gao, Kim,
  Lavernia, Mathaudhu, Mughrabi, Ritchie et~al.}]{zhu2021heterostructured}
\bibinfo{author}{Zhu, Y.}, \bibinfo{author}{Ameyama, K.},
  \bibinfo{author}{Anderson, P.M.}, \bibinfo{author}{Beyerlein, I.J.},
  \bibinfo{author}{Gao, H.}, \bibinfo{author}{Kim, H.S.},
  \bibinfo{author}{Lavernia, E.}, \bibinfo{author}{Mathaudhu, S.},
  \bibinfo{author}{Mughrabi, H.}, \bibinfo{author}{Ritchie, R.O.}, et~al.,
  \bibinfo{year}{2021}.
\newblock \bibinfo{title}{Heterostructured materials: superior properties from
  hetero-zone interaction}.
\newblock \bibinfo{journal}{Materials Research Letters} \bibinfo{volume}{9},
  \bibinfo{pages}{1--31}.
\newblock \DOIprefix\doi{https://doi.org/10.1080/21663831.2020.1796836}.
\bibitem[{Zhu and Wu(2023)}]{zhu2023heterostructured}
\bibinfo{author}{Zhu, Y.}, \bibinfo{author}{Wu, X.}, \bibinfo{year}{2023}.
\newblock \bibinfo{title}{Heterostructured materials}.
\newblock \bibinfo{journal}{Progress in Materials Science}
  \bibinfo{volume}{131}, \bibinfo{pages}{101019}.
\newblock \DOIprefix\doi{https://doi.org/10.1016/j.pmatsci.2022.101019}.

\end{thebibliography}

\end{document}